\documentclass[aps,pre,floatfix,twocolumn,nofootinbib]{revtex4}
\usepackage{graphicx} 
\usepackage{amsmath} 
\usepackage{amssymb}
\usepackage{subfigure}
\usepackage[usenames,dvipsnames,svgnames,table]{xcolor}
\newcommand{\comment}[1]{}
\newcommand{\BEQ}{\begin{equation}}
\newcommand{\EEQ}{\end{equation}}
\newcommand{\BEA}{\begin{eqnarray}}
\newcommand{\EEA}{\end{eqnarray}}
\renewcommand{\d}{{\rm d}}

\newcommand{\p}{\widetilde{p}}

\renewcommand{\P}{{\cal P}}
\newcommand{\Q}{{\cal Q}}

\newcommand{\la}{\lambda}

\begin{document}
\title{ Opinion Dynamics with  Confirmation Bias }

\author{Armen
  E. Allahverdyan$^{1)}$\footnote{armen.allahverdyan@gmail.com} 
and Aram Galstyan$^{2)}$}

\address{$^{1)}$
Department of Theoretical Physics, 
Yerevan Physics Institute, Alikhanian Brothers Street
2, Yerevan 375036, Armenia,\\
$^{2)}$
USC Information Sciences Institute, 4676 Admiralty Way, 
Marina del Rey, CA 90292, USA}

\begin{abstract} {\it Background.} Confirmation bias is the tendency
  to acquire or evaluate new information in a way that is consistent
  with one's preexisting beliefs. It is omnipresent in psychology,
  economics, and even scientific practices. Prior theoretical research
  of this phenomenon has mainly focused on its economic implications
  possibly missing its potential connections with broader notions of
  cognitive science.

    {\it Methodology/Principal Findings.}  We formulate a
    (non-Bayesian) model for revising subjective probabilistic opinion
    of a confirmationally-biased agent in the light of a persuasive
    opinion. The revision rule ensures that the agent does not react
    to persuasion that is either far from his current opinion or
    coincides with it. We demonstrate that the model accounts for the
    basic phenomenology of the social judgment theory, and allows to
    study various phenomena such as cognitive dissonance and boomerang
    effect. The model also displays the order of presentation
    effect|when consecutively exposed to two opinions, the preference
    is given to the last opinion (recency) or the first opinion
    (primacy)|and relates recency to confirmation bias. Finally, we
    study the model in the case of repeated persuasion and analyze its
    convergence properties.

    {\it Conclusions.} {The standard Bayesian approach to
      probabilistic opinion revision is inadequate for describing the
      observed phenomenology of persuasion process.} The simple
    non-Bayesian model proposed here does agree with this
    phenomenology and is capable of reproducing a spectrum of effects
    observed in psychology: primacy-recency phenomenon, boomerang
    effect and cognitive dissonance. We point out several limitations
    of the model that should motivate its future development.

\end{abstract}


\maketitle

\section*{\large Introduction}

{\it Confirmation bias} is the tendency to acquire or process new
information in a way that confirms one's preconceptions and avoids
contradiction with prior beliefs \cite{nickerson}. Various
manifestations of this bias have been reported in cognitive psychology
\cite{allakh,wason}, social psychology \cite{oskamp,darley_gross},
politics \cite{lazarsfeld} and (media) economics
\cite{rabin,yariv,mulla,shapiro}. Recent evidence suggests that
scientific practices too are susceptible to various forms of
confirmation bias \cite{nickerson,koehler,jeng,klayman,grif}, even
though the imperative of avoiding precisely this bias is frequently
presented as one of the pillars of the scientific method.

Here we are interested in the opinion revision of an agent $\P$ who is
persuaded (or advised) by another agent $\Q$
\cite{aronson,baron,nickerson}. (Below we use the terms {\em opinion}
and {\em belief} interchangeably.) We follow the known framework for
representing uncertain opinions of both agents via the subjective
probability theory \cite{baron}. Within this framework, the opinion of
an agent about propositions (events) is described by probabilities
that quantify his degree of confidence in the truth of these
propositions \cite{baron}. As we argue in the next section, the
standard Bayesian approach to opinion revision is inadequate for
describing persuasion. Instead, here we study confirmationally-biased
persuasion within the opinion combination approach developed in
statistics; see \cite{genest_zidek,clemen_winkler} for reviews.

We suggest a set of conditions that model cognitive aspects of
confirmation bias. Essentially, those conditions formalize the
intuition that the agent $\P$ does not change his opinion if the
persuasion is either far away or identical with his existing opinion
\cite{shreider,beim}. We then propose a simple opinion revision rule
that satisfies those conditions and is consistent with the ordinary
probability theory. The rule consists of two elementary operations:
{\em averaging} the initial opinion with the persuading opinion via
linear combination, and then {\em projecting} it onto the initial
opinion. The actual existence of these two operations has an
experimental support
\cite{anderson_review,anderson_book,yaniv_trimming,advising_review}.

We demonstrate that the proposed revision rule is consistent with the
{\it social judgment theory} \cite{aronson}, and reproduces the so
called {\it change-discrepancy} relationship
\cite{aronson,hovland,Whittaker1963,laroche,kaplo}. Furthermore, the
well-studied {\it weighted average} approach
\cite{anderson_book,fink_mink} for opinion revision is shown to be a
particular case of our model.

Our analysis of the revision rule also reveals novel effects. In
particular, it is shown that within the proposed approach, the recency
effect is related to confirmation bias. Also, repeated persuasions are
shown to hold certain monotonicity features, but do not obey the law
of diminishing returns. We also demonstrate that the rule reproduces
several basic features of the {\it cognitive dissonance} phenomenon
and predicts new scenarios of its emergence. Finally, the so called
{\it boomerang} ({\it backfire}) effect can emerge as an extreme form
of confirmation bias. The effect is given a straightforward
mathematical description in qualitative agreement with experiments.

The rest of this paper is organized as follows. In the next section we
introduce the problem setup and provide a brief survey of relevant
work, specifically focusing on inadequacy of the standard Bayesian
approach to opinion revision under persuasion. In the third section we
define our axioms and introduce the confirmationally biased opinion
revision rule. The fourth section \ref{social_gauss} relates our setup
to the social judgment theory. Next two sections describe how our
model accounts for two basic phenomena of experimental social
psychology: opinion change versus discrepancy and the order of
presentation effect. The seventh section shows how our model
formalizes features of cognitive dissonance, followed by analysis of
opinion change under repeated persuasion. Then we study the {\em
  boomerang effect}|the agent changes his opinion not towards the
persuasion, but against it| as a particular case of our approach. We
summarize and conclude in the last section.

\section*{\large The set-up and previous research}

Consider two agents $\P$ and $\Q$. They
  are given an uncertain quantity (random variable) ${\cal X}$ with
  values $k=1,...,N$, e.g. $k=({\it rain, ~ no rain})$, if this is a
  weather forecast. ${\cal X}$ constitutes the state of the world for
  $\P$ and $\Q$.  
The opinions of the agents are quantified via
  probabilities 
\BEA \label{cruz} p=\{p_k\}_{k=1}^N ~{\rm and}~
q=\{q_k\}_{k=1}^N, ~~ \sum_{k=1}^N p_k=\sum_{k=1}^N q_k=1, \EEA for
$\P$ and $\Q$ respectively.
 
Let us now assume that $\P$ is persuaded (or advised) by $\Q$.
(Persuasion and advising are not completely equivalent
\cite{yaniv_advice}. However, in the context of our discussion it will
be useful to employ both terms simultaneously stressing their commmon
aspects.) Throughout this paper we assume that the state of the world
does not change, and that the agents are aware of this fact. Hence,
$\P$ is going to change his opinion only under influence of the
opinion of $\Q$, and not due to any additional knowledge about ${\cal
  X}$ (For more details on this point see \cite{alc,keller} and the
second section of Supporting Information.)

{ The normative standard for opinion revision is related to
  the Bayesian approach. Below we discuss the main elements of the
  Bayesian approach, and outline certain limitations that motivates
  the non-Bayesian revision rule suggested in this work.

Within the Bayesian approach, the agent $\P$ treats his own
probabilistic opinion $p=\{p_k\}_{k=1}^N$ as a prior, and the
probabilistic opinion $q=\{q_k\}_{k=1}^N$ of $\Q$ as an evidence
\cite{genest_zidek,brown,french}. Next, it is assumed that $\P$ is
endowed with conditional probability densities $\Pi(q|k)$, which
statistically relate $q$ to the world state $k$. Upon receiving the
evidence from $\Q$, agent $\P$ modifies his opinion from $p_k$ to
$p(k|q)$ via the Bayes rule:
\begin{eqnarray}
  \label{baba}
p(k|q)=\Pi(q|k)p_k\left/\sum_{l=1}^N \Pi(q|l)p_l\right. .  
\end{eqnarray}
One issue with the Bayesian approach is that the assumption on the
existence and availability of $\Pi(q|k)$ may be too
strong~\cite{genest_zidek,diaconis_zabell,baron}. Another issue is
that existing empirical evidence suggests that people do not behave
according to the Bayesian approach~\cite{baron,amos}, e.g. they
demonstrate the order of presentation effect, which is generally
absent within the Bayesian framework.

In the context of persuasion, the Bayesian approach (\ref{baba}) has
two additional (and more serious) drawbacks. To explain the first
drawback, let us make a generic assumption that there is a unique
index $\hat{k}$ for which $\Pi(q|k)$ is maximized as a function of $k$
(for a given $q$): $\Pi(q|\hat{k})> \Pi(q|k)$ for $\hat{k}\not=k$.

Now consider repeated application of (\ref{baba}), which corresponds
to the usual practice of repeated persuasion under the same opinion
$q$ of $\Q$. The opinion of the agent then tends to be completely
polarized, i.e. $p_{\hat{k}}\to 1$ and $p_{k}\to 0$ for $k\not
=\hat{k}$. In the context of persuasion or advising, we would rather
expect that under repeated persuasion the opinion of $\P$ will
converge to that of $\Q$.

The second issue is that, according to (\ref{baba}), $\P$ will change
his opinion even if he has the same opinion as $\Q$: $p=q$.  {This
  feature may not be realistic: we do not expect $\P$ to change his}
opinion, if he is persuaded towards the same opinion he has
already. This drawback of (\ref{baba}) was noted in \cite{french}.
(Ref.~\cite{french} offers a modification of the Bayesian approach
that complies with this point, as shown in \cite{french} on one
particular example. However, that modification betrays the spirit of
the normative Bayesianism, because it makes conditional probabilities
depending on the prior probability.)

It is worthwhile to note that researchers have studied several aspects
of confirmation bias by looking at certain deviations from the Bayes
rule, e.g. when the conditional probability are available, but the
agent does not apply the proper Bayes rule deviating from it in
certain aspects \cite{rabin,yariv,mulla,shapiro}. One example of this
is when the (functional) form of the conditional probability is
changed depending on the evidence received or on the prior
probabilities. Another example is when the agent does not employ the
full available evidence and selects only the evidence that can
potentially confirm his prior expectations \cite{wason,lord,sugden}.
More generally, one has to differentiate between two aspects of the
confirmation bias that can be displayed either with respect to
information acquiring, or information assimilation (or both)
\cite{nickerson}. Our study will concentrate on information
assimilation aspect; first, because this aspect is not studied
sufficiently well, and second, because because it seems to be more
directly linked to cognitive limitations \cite{nickerson}. {We
  also stress that we focus on the belief revision, and not on actions
  an agent might perform based on those beliefs.}

\comment{I suggest to remove this paragraph, especially since the
  second reviewer asks so too.  Let us also discuss the difference
between {\it instrumental} and {\it expressive} opinions (beliefs)
\cite{abelson}. Certain opinions (beliefs) are {\it instrumental}, in
the sense that they lead to actions that serve definite material
goals~\cite{abelson}. However, there are also non-instrumental
opinions called {\it expressive}, which can lead to actions to which
one cannot attach any explicit utility function
\cite{abelson}. Expressive actions generally cannot be tested
empirically (at least within any reasonable time)
\cite{abelson}. Nevertheless, is has been established that
confirmation bias is developed {\it also} with respect to expressive
opinions \cite{nickerson}. }

\section*{\large Opinion revision rule}
\label{sec:model}

We propose the following conditions that the opinion revision rule
should satisfy.

{\bf 1.} The revised opinion $\p_k$ of $\P$ is represented as 
  \BEA
  \label{axiom1}
  \p_k= F[p_k, q_k]\left/{\sum}_{l=1}^N F[p_l, q_l]\right., \EEA  {
    where $F[x_1,x_2]$ is defined over $x_1\in [0,\infty)$ and $x_2\in
    [0,\infty)$. We enlarged the natural range $x_1\in [0,1]$ and
    $x_2\in [0,1]$, since below we plan to consider probabilities that
    are not necessarily normalized to $1$. There are at least two
    reasons for doing so: First, experimental studies of opinion elicitation
    and revision use more general
    normalizations \cite{anderson_review,anderson_book}. For example,
    if the probability is elicited in percents, the overall
    normalization is $100$. Second, and more importantly, the axioms
    defining subjective (or logical) probabilities leave the overall
    normalization as a free parameter \cite{cox}.

    We require that $F[x_1,x_2]$ is continuous for $x_1\in
    [0,\infty)$ and $x_2\in [0,\infty)$ and infinitely differentiable
    for $x_1\in (0,\infty)$ and $x_2\in (0,\infty)$. Such (or similar)
    conditions are needed for features that are established for
    certain limiting values of the arguments of $F$
    (cf. (\ref{axiom2}, \ref{axiom4})) to hold approximately whenever
    the arguments are close to those limiting values.} $F$ can also
  depend on model parameters, as seen below.

  Eq.~(\ref{axiom1}) means that $\P$ first evaluates the
  (non-normalized) weight $F[p_k, q_k]$ for the event $k$ based {\it solely}
  on the values  of $p_k$ and $q_k$, and then applies overall
  normalization.  { A related feature of (\ref{axiom1})
    is that it is local: assume that $N\geq 3$ and only  the
    probability $q_1$ is communicated by $\Q$ to $\P$. This suffices
    for $\P$ to revise his probability from $p_1$ to $\p_1$, and then
    adjust other probabilities via renormalization:
\begin{eqnarray}
&&  \p_1=F[p_1,q_1]\left/(\, F[p_1,q_1]+1-p_1)\right., \nonumber\\   
&&  \p_{k}=p_{k}\left/(\, F[p_1,q_1]+1-p_1)\right. ~~ k\geq 2.
  \label{ccc}
\end{eqnarray}
Eq.~(\ref{axiom1}) can be considered as a succession of such local
processes. 
}

{\bf 2.} If $p_k=0$ for some $k$, then $\p_k=0$: 
\begin{eqnarray}
  \label{axiom2}
F[0, y]=0.
\end{eqnarray}
 { The rationale of this condition is that if $\P$ sets the
  probability of a certain event strictly to zero, then he sees
  logical (or factual) reasons for prohibiting the occurrence of this
  event. Hence $\P$ is not going to change this zero probability under
  persuasion. } 


{\bf 3.} If $p_kq_k=0$ for all $k$, then $\p_k=p_k$: $\P$ cannot be
persuaded by $\Q$ if their opinions have no overlap.  

{\bf 4.} { If $\Q$'s and $\P$'s opinions are identical, then the
  latter will not change his opinion:
  $\{p_k\}_{k=1}^N=\{q_k\}_{k=1}^N$ (for all $k$) leads to
  $\{\p_k\}_{k=1}^N=\{p_k\}_{k=1}^N$. This can be written as
\begin{eqnarray}
  \label{axiom4}
  F[x, x]= x.
\end{eqnarray}
Conditions {\bf 3} and {\bf 4} are motivated by experimental
results in social psychology, which state that people are not persuaded
by opinions that are either very far, or very close to their initial
opinion \cite{Whittaker1963,Bochner1966,aronson}.

(Recall that we do not allow the uncertain quantity ${\cal X}$ to
change during the persuasion or advising. If such a change is allowed,
{\bf 4} may not be natural as the following example shows. Assume that
$\P$ holds a probabilistic opinion $(0.1,0.9)$ on a binary ${\cal X}$.
Let $\P$ learns that ${\cal X}$ changed, but he does not know in which
specific way it did. Now $\P$ meets $\Q$ who has the same opinion
$(0.1,0.9)$. Provided that $\Q$ does not echo the opinion of $\P$, the
agent $\P$ should perhaps change his opinion by decreasing the first
probability ($0.1$) towards a smaller value, because it is likely that
${\cal X}$ changed in that direction.)

{\bf 5.}  { $F$ is a homogeneous function
  of order one:}
\begin{eqnarray}
  \label{axiom5}
  F[\gamma x,\gamma y]=\gamma   F[ x,y] ~~~ {\rm
    for}~~~ \gamma\geq 0.
\end{eqnarray}
{ The rationale for this condition comes from the fact that
  (depending on the experimental situation) the subjective probability
  may be expressed not in normalization one (i.e. not with
  $\sum_{k=1}^N p_k=\sum_{k=1}^N q_k=1$), but with a different overall
  normalization (e.g. $\sum_{k=1}^N p_k=\sum_{k=1}^N q_k=\gamma$)
  \cite{anderson_review,anderson_book,cox}; cf. {\bf 1}.  In this
  light, (\ref{axiom5}) simply states that any choice of the overall
  normalization is consistent with the sought rule provided that it is
  the same for $\P$ and $\Q$. Any rescaling of the overall
  normalization by the factor $\gamma$ will rescale the non-normalized
  probability by the same factor $\gamma$; cf. (\ref{axiom5}).}


{\bf 6.} {Now we assume that the opinion assimilation by $\P$
  consists of two sub-processes. Both are related to heuristics of
  human judgement.

  {\bf 6.1} $\P$ combines his opinion {\em linearly} with the opinion
  of $\Q$~\cite{anderson_review,anderson_book,genest_zidek,
    genest_conway,advising_review}:
\begin{eqnarray}
  \label{axiom6.1}
  \widehat{p}_k=\epsilon p_k+(1-\epsilon)q_k, \qquad 0\leq \epsilon\leq 1,
\end{eqnarray}
where $\epsilon$ is a weight.  Several mathematical interpretations of
the weight $\epsilon$ were given in statistics, where (\ref{axiom6.1})
emerged as one of the basic rules of probabilistic opinion combination
\cite{genest_conway,bernardo}; see section I of Supporting Information.
One interpretation suggested by this approach is that
$\epsilon$ and $1-\epsilon$ are the probabilities|from the subjective
viewpoint of $\P$|for, respectively, $p$ and $q$ to be the true
description of states of the world \cite{genest_conway}: it is not
known to $\P$ which one of these probabilities ($p$ or $q$) conveys a
more accurate reflection of the world state. Then
$\{\widehat{p}_k\}_{k=1}^N$ is just the marginal probability for the
states of the world. There is also an alternative (normative) way of
deriving (\ref{axiom6.1}) from maximization of an average utility that
under certain natural assumptions can be shown to be the (negated)
average information loss \cite{bernardo}; see section I of 
Supporting Information for more details.

Several qualitative factors contribute to the subjective assessment of
$\epsilon$. For instance, one interpretation is to relate $\epsilon$
to credibility of $\Q$ (as perceived by $\P$): more credible $\Q$
leads to a larger $1-\epsilon$ \cite{advising_review}. Several other
factors might affect $\epsilon$: egocentric attitude of $\P$
that tends to discount opinions, simply because they do not belong to
him; or the fact that $\P$ has access to internal reasons for choosing
his opinion, while he is not aware of the internal reasons of $\Q$
{\it etc} \cite{advising_review}. Taking into account various factors
that contribute to the interpretation of $\epsilon$, we will treat it
as a free model parameter.

{\bf 6.2} Note that (\ref{axiom6.1}) does not satisfy conditions {\bf
  2} and {\bf 3} above.  We turn to the last ingredient of the sought
rule, which, in particular, should achieve consistency with conditions
{\bf 2} and {\bf 3}.

Toward this goal, we assume that $\P$ projects the linearly combined opinion $\widehat{p}$ (see
(\ref{axiom6.1})) onto his original opinion $p$. Owing to 
(\ref{axiom1}), we write this transformation as \BEA
  \label{axiom6.2}
  \p_k= \phi [p_k, \widehat{p}_k]\left/{\sum}_{l=1}^N 
\phi [p_l, \widehat{p}_l]\right. , \EEA 
where the function $\phi$ is to be determined. 

The above projection operation relates to {\it trimming}
\cite{yaniv_trimming,advising_review}, a human cognitive heuristics,
where $\P$ tends to neglect those aspects of $\Q$'s opinion  that
deviate from a certain reference. In the simplest case this reference
will be the existing opinion of $\P$.

To make the projection process (more) objective, we shall assume that
it commutes with the probabilistic revision: whenever
\begin{eqnarray}
  \label{upo}
  p_k'=\frac{ \gamma_kp_k}{\sum_{l=1}^N\gamma_lp_l}, ~~
  \widehat{p}_k'=\frac{ \gamma_k\widehat{p}_k}{\sum_{l=1}^N
   \gamma_l\widehat{p}_l}, 
  ~~ 1\leq k\leq N,
\end{eqnarray}
where $\gamma_k={\rm Pr}(...|k)>0$ are certain conditional probabilities,
$\p$ is revised via the same rule (\ref{upo}): 
\begin{eqnarray}
  \label{revo}
  \p'_k\equiv\frac{\phi [p_k', \widehat{p}_k']}{{\sum}_{l=1}^N 
    \phi [p_l', \widehat{p}_l']}
=\frac{\gamma_k\phi [p_k, \widehat{p}_k]}{{\sum}_{l=1}^N 
   \gamma_l \phi [p_l, \widehat{p}_l]}.
\end{eqnarray}
This feature means that the projection is consistent with probability
theory: it does not matter whether (\ref{axiom1}) is applied before or
after (\ref{upo}).

It is known that (\ref{axiom6.2}) together with 
(\ref{upo}, \ref{revo}) selects a unique function
\cite{genest_zidek}:
\begin{eqnarray}
  \label{berd}
  \phi [p_k, \widehat{p}_k]=p_k^{\mu}\,\, \widehat{p}_k^{1-\mu}, 
~~ 0< \mu \leq 1
\end{eqnarray}
where $\mu$ quantifies the projection strength: for $\mu=1$ the
projection is so strong that $\P$ does not change his opinion at all
(conservatism), while for $\mu\to 0$, $\P$ fully accepts $\widehat{p}$
(provided that $p_k>0$ for all $k$). (The above commutativity is
formally valid also for $\mu\leq 0$ or $\mu>1$, but both these cases
are in conflict with (\ref{axiom2}).) In particular, $\epsilon\to 0$
and $\mu\to 0$ is a limiting case of a fully credulous agent that
blindly follows persuasion provided that all his probabilities are
non-zero. (For a sufficiently small $\mu$, a small $p_k$ is less
effective in decreasing the final probability $\tilde{p}_k$; see
(\ref{berd}). This is because $p_k^\mu=e^{\mu\ln p_k}$ tends to zero
for a fixed $\mu$ and $p_k\to 0$, while it tends to one for a fixed
$p_k$ and $\mu\to 0$. This interpaly between $p_k\to 0$ and $\mu\to 0$
is not unnatural, since the initial opinion of a credulous agent is
expected to be less relevant.}}. {The case of credulous agent is of an
intrinsic interest and it does warrant further studies. However, since
our main focus is confirmation bias, below we set $\mu=1/2$ and
analyze the opinion dynamics for varying $\epsilon$.)

The final opinion revision rule reads from (\ref{berd},
\ref{axiom6.1}, \ref{axiom6.2}): \BEA
\label{3}
\p_k=\frac{\sqrt{p_k [\epsilon p_k+ (1-\epsilon)q_k ]}}{{\sum}_{l=1}^N
  \sqrt{p_l [\epsilon p_l+ (1-\epsilon)q_l ]}}, \qquad 0\leq
\epsilon\leq 1. 
\EEA 
It is seen to satisfy conditions {\bf 1-5}. 

(Note that the analogue of (\ref{revo}), $p_k'\propto{ \gamma_kp_k}$,
$q_k'\propto{ \gamma_kq_k}$ does not leave invariant the linear
function (\ref{axiom6.1}). First averaging, $\epsilon
p_k+(1-\epsilon)q_k$ and then applying $p_k'\propto{ \gamma_kp_k}$,
$q_k'\propto{ \gamma_kq_k}$ is equivalent to first applying the latter
rules and then averaging with a different weight $\epsilon'$.  This is
natural: once $\epsilon$ can be (in principle) interpreted as a
probability it should also change under probabilistic revision
process.)

The two processes were applied above in the specific order:
first averaging (\ref{axiom6.1}), and then projection
(\ref{axiom6.2}). We do not have any strong objective justifications for
 this order, although certain experiments on advising indicate on
the order that led to (\ref{3}) \cite{yaniv_trimming}. Thus, it is not
excluded that the two sub-processes can be applied in the reverse
order: first projection and then averaging. Then instead of (\ref{3})
we get (\ref{axiom1}) with: \BEA
\label{alter}
F[p_k,q_k;\epsilon,\mu]=\epsilon p_k+ (1-\epsilon)p_k^{\mu}
q_k^{1-\mu}, ~~ 0<\mu<1.  \EEA 
Our analysis indicates that
both revision rules (\ref{3}) and (\ref{alter}) (taken with $\mu=1/2$)
produce qualitatively similar results. Hence, we focus on (\ref{3})
for the remainder of this paper.

Returning to (\ref{cruz}), we note that $k=x$ can be a continuous
variable, if (for example) the forecast concerns the chance of having
rain or the amount of rain. Then the respective probability densities
are:
\begin{eqnarray}
  \label{dens}
  p(x) {\rm ~~and~~} q(x), ~~~ 
\int\d x\, p(x)=\int\d x\, q(x)=1.  
\end{eqnarray}
 {Since the revision rule (\ref{3}) is continuous and
  differentiable (in the sense defined after (\ref{axiom1})), it
  supports a smooth transition between discrete probabilities and
  continuous and differentiable probability densities}. In
particular, (\ref{3}) can be written directly for densities: for
$p_k\simeq p(x_k)\d x$ we obtain from (\ref{3})
\begin{eqnarray}
  \label{30}
  \p(x)=
  \frac{\sqrt{\, p(x) [\,\epsilon p(x)+ (1-\epsilon)q(x)\, ]\, }}
{\int\d x'\,
    \sqrt{\, p(x') [\,\epsilon p(x')+ (1-\epsilon)\,q(x')\, ]\,}}.
\end{eqnarray}
}

\section*{\large Social judgment theory and Gaussian opinions}
\label{social_gauss}

\subsection*{Opinion latitudes}

Here we discuss our model in the context of the social judgment theory
\cite{handbook,aronson}, and consider several basic scenarios of
opinion change under the rule (\ref{30}).

According to the social judgment theory, an agent who is exposed to
persuasion perceives and evaluates the presented information by
comparing it with his existing attitudes (opinions). The theory
further postulates that an attitude is composed of three zones, or
latitudes: {\it acceptance}, {\it non-commitment} and {\it rejection}
\cite{handbook,aronson}.  { The opinion that is most acceptable to
  $\P$,} or the {\it anchor}, is located at the center of the latitude
of acceptance. The theory states that persuasion does not change the
opinion much, if the persuasive message is either very close to the
anchor or falls within the latitude of rejection
\cite{handbook,aronson}. The social judgment theory is popular, but
its quantitative modeling has been rather scarce.  In particular, to
the best of our knowledge, there has been no attempt to develop a
consistent probabilistic framework for the theory.
(The literature on the social judgment theory offers some formal
mathematical expressions that could be fitted to experimental data
\cite{laroche}. There is also a more quantitative theory
\cite{Hogarth1992} whose content is briefly reminded in section III 
of Supporting Information.)

Let us assume that $k=x$ is a continuous variable (cf. (\ref{dens}))
and that $p(x)$ and $q(x)$ are Gaussian with mean $m_\lambda$ and
dispersion $v_\lambda$ ($\lambda=\P,\Q$): \BEA
\label{38}
p(x)=\frac{e^{-\frac{(x-m_\P)^2}{2v_\P}}}{\sqrt{2\pi v_\P}} , \qquad
q(x)=\frac{e^{-\frac{(x-m_\Q)^2}{2v_\Q}}}{\sqrt{2\pi v_\Q}}.  \EEA
 { Effectively, Gaussian probabilistic opinions are produced in
  experiments, when the subjects are asked to generate an opinion with
  $\approx 95 \%$ confidence in a certain interval
  \cite{advising_review}.} Now we can identify the anchor with the
most probable opinion $m_\lambda$, while $v^{-1}_\lambda$ quantifies
the opinion uncertainty.

The latitude of acceptance amounts to opinions not far from the
anchor, while the latitude of rejection contains close-to-zero
probability events, since $\P$ does not change his opinion on them;
recall point {\bf 2} from the previous section.
One can also identify the three latitudes with appropriately chosen
zones in the distribution. For instance, it is plausible to define
the latitudes of acceptance and rejection by, respectively, the
following formulas of the $3\sigma$ rule known in statistics
\begin{gather}
  \label{lat}
x\in[m_\P-2\sqrt{v_\P}, \, \, \,  m_\la+2\sqrt{v_\P}],\\  
  \label{lat_r}
x\in (-\infty, m_\P-3\sqrt{v_\P}]\cup [m_\P+3\sqrt{v_\P},\infty),
\end{gather}
where the latitude of non-commitment contains whatever is left out
from (\ref{lat}, \ref{lat_r}). Recall that the
latitudes of acceptance, non-commitment and rejection carry
(respectively) 95.4, 4.3 and 0.3 \% of probability.

While the definitions (\ref{lat}, \ref{lat_r}) are to some extent
arbitrary, they work well with the rule (\ref{30}), e.g. if the
opinions of $\P$ and $\Q$ overlap only within their rejection
latitudes, then neither of them can effectively change the opinion of
another. Also, $\P$ is persuaded most strongly, if the anchor of the
persuasion falls into the non-commitment latitude of $\P$. This is
seen below when studying change-discrepancy relations.

\subsection*{Weighted average of anchors}

Next, we demonstrate that the main quantitative theory of persuasion
and opinion change|the weighted average approach
\cite{anderson_book,fink_mink}|is a particular case of our model. 
 { We assume that the opinions $p(x)$ and $q(x)$ are given as 
  \begin{eqnarray}
    \label{bora}
    p(x)=f(x-m_\P), \qquad     p(x)=g(x-m_\Q), \\
    f'(0)\equiv \d f/\d x|_{x=0}=g'(0)=0, \nonumber\\
    f''(0)<0, \qquad g''(0)<0,
  \end{eqnarray}
  where both $f(x)$ and $g(x)$ have a unique maximum at $x=0$. Hence
  $p(x)$ (resp. $q(x)$) has a single anchor (maximally probable
  opinion) $m_\P$ (resp. $m_\Q$); see (\ref{38}) for concrete
  examples.

  If $|m_\P-m_\Q |$ is sufficiently small, $\p(x)$ given by
  (\ref{bora}, \ref{30}) has a single anchor which is shifted towards
  that of $q(x)$; see Fig.~\ref{fig_1_a}.  We now look for the maximum
  $m_{\tilde{\P}}$ of $\p(x)$ by using (\ref{bora}) in (\ref{30}). We
  neglect factors of order ${\cal O}[ (m_\P-m_\Q )^2/v_\P]$ and ${\cal
    O}[ (m_\P-m_\Q )^2/v_\Q]$ and deduce:}
\begin{eqnarray}
  \label{lin}
  &&  m_{\tilde{\P}}=(1-\alpha_\Q)\, m_\P + \alpha_\Q\, m_\Q ,\\
  &&  \alpha_{\Q}\equiv\frac{(1-\epsilon)\frac{|g''(0)|}{g(0)}}
  {(1-\epsilon)\frac{|g''(0)|}{g(0)}+
    \frac{|f''(0)|}{f(0)}\left[2\epsilon\frac{f(0)}{g(0)}+1-\epsilon\right] 
  }.
  \end{eqnarray}
  Eq.~(\ref{lin}) is the main postulate of the weighted average
  approach; see \cite{anderson_book,fink_mink} for reviews. Here
  $\alpha_\Q$ and $1-\alpha_\Q$ are the weights of $\Q$ and of
  $\P$, respectively. For the Gaussian case (\ref{38}), we have
\begin{eqnarray}
  \label{lino}
\alpha_\Q=\left[ 1+y\left( 1+\frac{2\epsilon \sqrt{y}}{1-\epsilon} 
\right)  \right]^{-1}, \qquad  y\equiv \frac{v_\Q}{v_\P}.
  \label{linux}
\end{eqnarray}
Furthermore, we have  
\begin{eqnarray}
 \frac{\partial \alpha_\Q}{\partial y}= -\alpha_\Q^2\left[  
1+ \frac{3\epsilon \sqrt{y}}{1-\epsilon} 
\right], ~~~~
   \frac{\partial \alpha_\Q}{\partial \epsilon}= -
\frac{2 y^{3/2}\, \alpha_\Q^2}{(1-\epsilon)^2},
\end{eqnarray}
Thus, $\alpha_\Q$'s dependence on the involved parameters is
intuitively correct: it increases with the confidence $1/v_\Q$ of
$\Q$, and decreases with the confidence $1/v_\P$ of $\P$. Note also
that $\alpha_\Q$ decreases with $\epsilon$.

\comment{ \footnote{Eq.~(\ref{lino}) is specific to rule (\ref{30}),
    e.g. if we apply rule (\ref{alter}) with $\mu=1/2$, then
    $\sqrt{\frac{v_\Q}{v_\P}}$ in (\ref{lino}) changes to
    $[\frac{v_\Q}{v_\P}]^{1/4}$. But the qualitative form of this
    dependence remains the same. }}

Now let $p(x)$ and $q(x)$ (and hence $\p(x)$) have the same maximum
$m_\P=m_\Q$, but $v_\P\approx v_\Q$; see (\ref{38}). Expanding
(\ref{30}, \ref{38}) over $v_\P- v_\Q$ and keeping the first-order
term only we get
\begin{eqnarray}
  \label{klim}
  v_{\widetilde{\P}}=\frac{1-\epsilon}{2}\,v_{\Q}+\frac{1+\epsilon}{2}\,v_{\P},
\end{eqnarray}
where $v_{\widetilde{\P}}$ is the dispersion of (non-Gaussian)
$\p(x)$. Eq.~(\ref{klim}) implies 
\begin{eqnarray}
  \label{sakhalin}
  ( v_{\widetilde{\P}}-  v_\P)(v_\Q - v_\P)=\frac{1-\epsilon}{2}\,
  (v_\Q - v_\P)^2 \geq 0,
\end{eqnarray}
i.e. if $1/v_\Q>1/v_\P$ (resp. $1/v_\Q<1/v_\P$), the final opinion of
$\P$ becomes more (resp. less) narrow than his initial
opinion. Fig.~\ref{fig_1_b} shows that $ ( v_{\widetilde{\P}}-
v_\P)(v_\Q - v_\P)\geq 0$ holds more generally.

Thus, the weighted average approach is a particular case of our model,
where the agent $\P$ is persuaded by a slightly different opinion.
Note also that our model suggests a parameter structure of the
weighted average approach.

\subsection*{Opinions and bump-densities}

{ Gaussian densities (with three latitudes) do correspond to the
  phenomenology of social psychology. However, in certain scenarios
  one might need other forms of densities, e.g., when the probability
  is strictly zero outside of a finite support. Such opinions can be
  represented by bump-functions
\begin{eqnarray}
  \label{bump}
  \chi(x;b)&=& {\cal N}(b)\exp[\frac{b}{x^2-1}] ~~ {\rm for}~~|x|< 1\\
           &=& 0 ~~~~ {\rm for}~~x\leq -1~~~{\rm and}~~~ x\geq 1.\nonumber
\end{eqnarray}
where $b>0$ is a parameter, ${\cal N}(b)$ is the normalization and the
support of the bump function was chosen to be $[-1,1]$ for
concretness. The advantage of the bump function that is infinitely
differentiable despite of having a finite support.

For sufficiently large $b$, $\chi(x;b)$ is close to a Gaussian, while for
small $b$, $\chi(x;b)$ represents an opinion that is (nearly)
homogeneous on the interval $[-1,1]$; see Fig.~\ref{fig_1_1}. The
opinion revision with bump densities follows to the general intuition
of rule (\ref{30}); see Fig.~\ref{fig_1_1}.  }

\section*{\large Opinion change vs discrepancy}
\label{change_d}

One of extensively studied questions in social psychology is how the
opinion change is related to the discrepancy between the initial
opinion and the position conveyed by the persuasive message
\cite{aronson,hovland,Whittaker1963,laroche,kaplo}. Initial studies
suggested a linear relationship between discrepancy and the opinion
change \cite{hovland}, which agreed with the prediction of the
weighted average model. Indeed, (\ref{lin}) yields the following
linear relationship between the change in the anchor and the initial
opinion discrepancy of $\P$ and $\Q$: \BEA
m_{\tilde{\P}}-m_{\P}=\alpha_\Q\, (m_{\Q}-m_{\P}).
\label{pushi}
\EEA 
However, consequent experiments revealed that the linear regime
is restricted to small discrepancies only and that the actual behavior
of the opinion change as a function of the discrepancy is
non-monotonic: the opinion change reaches its maximal value at some
discrepancy and decreases
afterward~\cite{aronson,Whittaker1963,laroche,kaplo}.

To address this issue within our model, we need to define
distance $h[p,q]$ between two probability densities $p(x)$ and
$q(x)$. Several such distances are known and standardly employed
\cite{gibbs}. Here we select the Hellinger distance (metric)
\begin{eqnarray}
  \label{hellinger}
 h[p,q] &\equiv& \frac{1}{\sqrt{2}}
\left[\int\d x[\, \sqrt{p(x)}-\sqrt{q(x)}\,]^2 \right]^{1/2}, \\
&=&\left[1-\int\d x\sqrt{p(x)q(x)} \right]^{1/2}.
\label{hellinger0}
  \end{eqnarray}
   { Since $\sqrt{p(x)}$ is a unit vector in the $\ell_2$ norm, 
    Eq.~(\ref{hellinger}) relates to the Euclidean ($\ell_2$-norm)
    distance.} It is applicable to discrete probabilities by changing
  the integral in (\ref{hellinger}, \ref{hellinger0}) to sum. For
  Gaussian opinions (\ref{38}) we obtain
 \begin{eqnarray}
  h[p,q] = \left[1-\left[\frac{(v_\Q v_\P)^{1/2}}
{\frac{v_\Q +v_\P}{2}}\right]^{1/2}
\,e^{ -\frac{(m_\Q -m_\P)^2}{4(v_\Q +v_\P)} }   \right]^{1/2}.
 \label{hellinger_g}
 \end{eqnarray}
 A virtue of the Hellinger distance is that it is a measure of overlap
 between the two densities; see (\ref{hellinger0}).  { We stress,
   however, that there are other well-known distances measures in
   statistics \cite{gibbs}. All results obtained below via the
   Hellinger distance will be checked with one additional metric, the
   total variation ($\ell_1$-norm distance):
 \begin{eqnarray}
   \label{totality}
   \delta[p,q]= \frac{1}{2}\int\d x\left|\, p(x)-q(x)\,\right|.
 \end{eqnarray}
}

(To motivate the choice of (\ref{totality}), let us recall two
important variational features of this distance \cite{gibbs}: (1)
$\delta[p,q] ={\rm max}_{\Omega\in R^1}\left|\int_\Omega \d
  x(p(x)-q(x))\,\right|$. (2) Define two (generally dependent) random
variables $X,Y$ with joint probability density $g(x,y)$ such that
$\int\d x\, g(x,y)=q(y)$, $\int\d y\, g(x,y)=p(x)$. Now it holds that
$\delta[p,q] ={\rm min}\left[ {\rm Pr}(X\not=Y) \right]$, where ${\rm
  Pr}(X\not=Y)=1-\int\d x\, g(x,x)$, and the minimization is taken
over all $g(x,y)$ with fixed marginals equal to $p(x)$ and $q(y)$,
respectively.)

The opinion change is characterized by the Hellinger distance
$h[p,\p]$ between the initial and final opinion of $\P$, while the
discrepancy is quantified by the Hellinger distance $h[p,q]$ between
the initial opinion of $\P$ and the persuading opinion. For
concreteness we assume that the opinion strengths $1/v_\P$ and
$1/v_\Q$ are fixed. Then $h[p,q]$ reduces to the distance $m=|m_\P
-m_\Q|$ between the anchors (peaks of $p(x)$ and $q(x)$); see
(\ref{hellinger_g}).

 Fig.~\ref{fig_2_a} shows that the change $h[p,\p]$ is maximal at 
 $m=m_{c\,h}$; it decreases for $m>m_{c\,h}$, since the densities of $\P$
 and $\Q$ have a smaller overlap. The same behavior is shown by the
 total variation $\delta[p,\p]$ that maximizes at  $m=m_{c\,\delta}$;
 see Fig.~\ref{fig_2_a}.

 The dependence of $m_{c\,h}$ (and of $m_{c\,\delta}$) on $\epsilon$
 is also non-monotonic; Fig.~\ref{fig_2_b}. This is a new prediction
 of the model.  Also, $m_{c\,h}$ and $m_{c\,\delta}$ are located
 within the latitude of non-commitment of $\P$ (this statement does
 not apply to $m_{c\,h}$, when $\epsilon$ is close to $1$ or $0$);
 cf. (\ref{lat}, \ref{lat_r}).  This point agrees with experiments
 \cite{aronson,Whittaker1963}.

 {
 Note that experiments in social psychology are typically carried out
 by asking the subjects to express one preferred opinion under
 given experimental conditions
 \cite{aronson,hovland,Whittaker1963,laroche,kaplo}. It is this single
 opinion that is supposed to change under persuasion. It seems
 reasonable to relate this single opinion to the maximally probable one 
 (anchor) in the probabilistic set-up. Thus, in addition to
 calculating distances, we show in Fig.~\ref{fig_2_c} how the final
 anchor $m_{\widetilde{\P}}$ of $\P$ deviates from his initial anchor
 $m_\P$.

 Fig.~\ref{fig_2_c} shows that for $\epsilon>0.25$, the behavior of
 $\Delta m=|m_{\widetilde{\P}}-m_\P|$ as a function of $m=|m_\P-m_\Q|$
 has an inverted-U shape, as expected. It is seen that $\Delta m$
 saturates to zero much faster compared to the distance $h[p,\p]$. In
 other words, the full probability $\p$ keeps changing even when the
 anchor does not show any change; cf. Fig.~\ref{fig_2_c} with
 Fig.~\ref{fig_2_a}.

A curious phenomenon occurs for a sufficiently small $\epsilon$; see
Fig.~\ref{fig_2_c} with $\epsilon=0.1$. Here $\Delta m$ drops suddenly
to a small value when $m$ passes certain crticial point;
Fig.~\ref{fig_2_c}. The mechanism behind this sudden change is as
follows: when the main peak of $p(x)$ shifts towards $m_\Q$, a second,
sub-dominant peak of $\p(x)$ appears at a value smaller than
$m_\P$. This second peak grows with $m$ and at some critical value it
overcomes the first peak, leading to a bistability region and an
abrupt change of $\Delta m$. The latter arises due to a subtle
interplay between the high credibility of $\Q$ (as expressed by a
relatively small value of $\epsilon$) and sufficiently large
discrepancy between $\P$ and $\Q$ (as expressed by a relatively large
value of $m$).  Recall, however, that the distance $h[p,\p]$
calculated via the full probability does not show any abrupt change.

The abrupt change of $\Delta m$ is widely discussed (and
experimentally confirmed) in the attitude change literature; see
\cite{cata} for a recent review. There the control variables for the
attitude change|information and involvement \cite{cata}|differ from
$\epsilon$ and $m$. However, one notes that the weight $\epsilon$ can
be related to the involvement: more $\P$ is involved into his existing
attitude, larger is $\epsilon$, while the discrepancy $m$ connects to
the (new) information contained in the persuasion ($m=0$ naturally
means zero information). }

Let us finally consider a scenario where the change-discrepancy
relationship is monotonic. It is realized for $m_\P=m_\Q$ (coinciding
anchors), where the distance (\ref{hellinger_g}) between $p(x)$ and
$q(x)$ is controlled by $v_\Q$ (for a fixed $v_\P$). In this case,
vthe change $h[p,\p]$ is a monotonic function of discrepancy $h[p,q]$:
a larger discrepancy produces larger change. This example is
interesting, but we are not aware of experiments that have studied the
change-discrepancy relation in the case of two identical anchors.

\section*{\large Order of presentation}
\label{section_order_presentation}

\subsection*{Recency versus primacy}

{When an agent is consecutively presented with two persuasive
  opinions, his final opinion is sensitive to the order of
  presentation
  \cite{diaconis_zabell,aronson,hovland,miller,Hogarth1992,baron,
    nickerson}. While the existence of this effect is largely
  established}, its direction is a more convoluted matter. (Note that
the order of presentation effect is not predicted by the Bayesian
approach; see (\ref{baba}).) Some studies suggest that the first
opinion matters more (primacy effect), whereas other studies advocate
that the last interaction is more important (recency effect). While it
is not completely clear which experimentally (un)controlled factors
are responsible for primacy and recency, there is a widespread
tendency of relating the primacy effect to confirmation bias
\cite{baron,nickerson}. This relation involves a qualitative argument
that we scrutinize below.

We now define the order of presentation effect in our situation. The
agent $\P$ interacts first with $\Q$ (with probability density
$q(x)$), then with $\Q'$ with probability density $q'(x)$. To
ensure that we compare {\it only} the order of $\Q$ and $\Q'$ and not
different magnitudes of influences coming from them, we take both
interactions to have the same parameter $0<\epsilon<1$. Moreover, we
make $\Q$ and $\Q'$ symmetric with respect to each other and with
respect to $\P$, e.g. if $p(x)$, $q(x)$ and $q'(x)$ are given by
(\ref{38}) we assume
\begin{eqnarray}
  \label{symo}
v_{\Q'}=v_{\Q}, ~~~ m_{\Q'}-m_{\P}=m_{\P}-m_{\Q}.    
\end{eqnarray}
We would like to know whether the final opinion $p(x|q,q')$ of $\P$ is 
closer to $q(x)$ (primacy) or to $q'(x)$ (recency).

In the present model (and for $0\leq\epsilon < 1$), the final opinion
$p(x|q,q')$ is {\em always} closer to the last opinion $q'(x)$, both
in terms of maximally probable value and distance. In other words, the
model unequivocally predicts the recency effect. In terms of the
Hellinger distance (\ref{hellinger})
\begin{eqnarray}
  \label{order_presentation} 
h[p(x|q,q'), q']<h[p(x|q,q'),q].
\end{eqnarray}
See Fig.~\ref{fig_3_a} for an example (In our model primacy effect
exists in the boomerang regime $\epsilon>1$; see below.)
 
To illustrate (\ref{order_presentation}) analytically on a specific
example, consider the following (binary) probabilistic opinion of
$\P$, $\Q$ and $\Q'$
\begin{eqnarray}
  \label{ord}
p=({1}/{2},\, {1}/{2}), ~~ q=(0,1), ~~ q'=(1,0). 
\end{eqnarray}
$\P$ is completely ignorant about the value of the binary  variable,
while $\Q$ and $\Q'$ are fully convinced in their opposite beliefs. 
If $\P$ interacts first with $\Q$ and then with $\Q'$ (both
interactions are given by (\ref{3}) with $\epsilon=\frac{1}{2}$), the
opinion of $\P$ becomes $(0.52727,\, 0.47273)$. This is closer to the
last opinion (that of $\Q'$).

The predicted recency effect in our model seems rather
counterintuitive. Indeed, since the first interaction shifts the
opinion of $\P$ towards that of $\Q$, one would think that the second
interaction with $\Q'$ should influences $\P$'s opinion less, due to a
smaller overlap between the opinions of $\Q'$ and $\P$ before the
second interaction. In fact, this is the standard argument that
relates primacy effect to the confirmation bias
\cite{nickerson,baron}: the first interaction shapes the opinion of
$\P$ and makes him confirmationally biased against the second
opinion. This argument does not apply to the present model due to the
following reason: even though the first interaction shifts $\P$'s
anchor towards $\Q$'s opinion, it also deforms the shape of the
opinion; see Fig.~\ref{fig_1_a}. And the deformation produced by our
revision rule happens to favor the second interaction more.

To get a deeper understanding of the recency effect, let us expand
(\ref{3}) for small $\eta\equiv 1-\epsilon$:
\BEA
\p_k=p_k+\frac{\eta}{2}(q_k-p_k)+\frac{\eta^2}{8}(p_k-1)\sum_l
\frac{(q_l-p_l)^2}{p_l} \nonumber\\
+ {\cal O}[\eta^3].
\label{c1}
\EEA
If now $\P$ interacts with an agent $\Q'$ having opinion $q'$, the
resulting opinion $p(q,q')$ reads from (\ref{c1}):
\BEA
\label{c2}
p_k(q,q')&&=p_k \nonumber\\
&&+\frac{\eta}{2}(q_k-p_k)+\frac{\eta^2}{8}(p_k-1)\sum_l
\frac{(q_l-p_l)^2}{p_l} \nonumber\\
&&+\frac{\eta}{2}(q'_k-p_k)+\frac{\eta^2}{8}(p_k-1)\sum_l
\frac{(q'_l-p_l)^2}{p_l}\nonumber\\
&&+\frac{\eta^2}{4}(p_k-q_k)
+ {\cal O}[\eta^3].
\EEA
Hence in this limit $p_k(q,q')-p_k(q',q)$
depends only on $q'_k-q_k$ (and not e.g. on $q_{l\not= k}$):
\BEA
p_k(q,q')-p_k(q',q)=\eta^2 [q'_k-q_k]/4+{\cal O}[\eta^3].
\label{gilgamesh}
\EEA It is seen that the more probable persuasive opinion (e.g.  the
opinion of $\Q'$ if $q'_k>q_k$) changes the opinion of $\P$ if it
comes later. This implies the recency effect. Indeed, due to symmetry
conditions for checking the order of presentation effect we can also
look at $h[p(q,q'),q]-h[p(q',q),q]$. Using (\ref{gilgamesh}) we get
for this quantity: $\frac{\eta^2}{16h[p(q',q),q]
}\sum_k[q_k-q'_k]\sqrt{q_k/p_k}>0$, again due to symmetry conditions.

  Note that this argument on recency directly extends to more general
  situations, where the agent is exposed to different opinions
  multiple times. For instance, consider an exposure sequence
  $q\,q\,q'\,q'$ and its reverse $q'\,q'\,q\,q$. It can be shown that
  the model predicts a recency effect in this scenario as well. For
  this case, we get instead of (\ref{gilgamesh}):
  $p_k(q,q')-p_k(q',q)=\eta^2 [q'_k-q_k]+{\cal O}[\eta^3]$.

Note that the primacy-recency effect is only one (though important!)
instance of contextual and non-commutative phenomena in psychology;
see \cite{atman,buse} and references therein. Hence in section IV of
Supporting Information we study a related (though somewhat less
interesting) order of presentation effect, while below we discuss our
findings in the context of experimental results.

\subsection*{Experimental studies of order of presentation effect}
\label{expo} 

{ We now discuss our findings in this section in
  the context of experimental results on primacy and recency. The
  latter can be roughly divided into several group: persuasion tasks
  \cite{aronson,miller}, symbol recalling \cite{wright}, inference
  tasks \cite{Hogarth1992}, and impression formation
  \cite{anderson,anderson_book}. In all those situations one generally
  observes both primacy and recency, though in different proportions
  and under different conditions \cite{Hogarth1992}.  Generally, the
  recency effect is observed whenever the {\em retention} time (the
  time between the last stimulus and the data taking) is short. If
  this time is sufficiently long, however, the recency effect changes
  to the primacy effect \cite{aronson,miller,stewart,wright}. The
  general interpretation of these results is that there are two
  different processes involved, which operate on different
  time-scales. These processes can be conventionally related to
  short-term and long-term memory \cite{wright}, with the primacy
  effect related to the long-term memory. In our model the longer time
  process is absent. Hence, it is natural that we see only the recency
  effect. The prevalence of recency effects is also seen in inference
  tasks, where the analogue of the short retention time is the
  incremental (step-by-step) opinion revision strategy
  \cite{Hogarth1992}.
  
  At this point, let us remind the importance of symmetry conditions
  [such as (\ref{symo})] for observing a {\it genuine} order of
  presentation effect. Indeed, several experimental studies|in
  particular those on impression formation|suggest that the order of
  presentation exists {\it due} to different conditions in the first
  versus the second interaction
  \cite{anderson,aronson,krug,Hogarth1992}.  (In our context, this
  means different parameters $\epsilon$ and $\epsilon'$ for each
  interaction). For instance, Refs.~\cite{anderson,aronson} argue that
  the primacy effect is frequently caused by attention decrement (the
  first action/interaction gets more attention); see also \cite{krug}
  in this context. This effect is trivially described by our model, if
  we assume $\epsilon$ to be sufficiently smaller than $\epsilon'$. In
  related experiments, it was shown that if the attention devoted to
  two interactions is balanced, the recency effect results
  \cite{hendrick}, which is consistent with the prediction of our
  model.

  At the same time, in another interesting study based on subjective
  probability revision, where the authors had taken special measures
  for minimizing the attention decrement, the results indicated a
  primacy effect \cite{peterson}.  }

We close this section by underlining the advantages and drawbacks of
the present model concerning the primacy-recency effect: the main
advantage is that it demonstrates the recency effect and shows that
the well-known argument on relating confirmation bias to primacy does
not hold generally. The main drawback is that the model does not
involve processes that are supposedly responsible for the
experimentally observed interplay between recency and primacy. In the
concluding section we discuss possible extensions of the model that
can account for this interplay.

\section*{\large Cognitive dissonance}
\label{cog_disso}

Consider an agent whose opinion probability density has two peaks on
widely separated events. Such a density|with the most probable opinion
being different from the average|is indicative of cognitive
dissonance, where the agent believes in mutually conflicting things
\cite{aronson,disso}. 


The main qualitative scenario for the emergence of cognitive
dissonance is when an agent|who initially holds a probabilistic
opinion with a single peak|is exposed to a conflicting information
coming from a sufficiently credible source \cite{aronson,disso}.
We now describe this scenario quantitatively. 

Consider again the opinion revision model (\ref{30}, \ref{38}), and
assume that $|m_\P-m_\Q |$ is neither very large nor very small (in
both these cases no serious opinion change is expected), $v_\Q/v_\P<1$
(self-assured persuasive opinion) and $0<\epsilon< 1$. In this case,
we get two peaks (anchors) for the final density $\p(x)$. The first
peak is very close to the initial anchor of $p(x)$, while the second
closer to the anchor of $q(x)$; see Fig.~\ref{fig_6_a_a}. Thus,
persuasion from $\Q$ whose opinion is sufficiently narrow and is
centered sufficiently close (but not too close) to $\P$'s initial
anchor leads to cognitive dissonance: $\P$ holds simultaneously two
different anchors, the old one and the one induced by $\Q$.

There are 3 options for reducing cognitive dissonance: 

 {
{\it (i)} Increase $\epsilon$ making it closer to $1$, i.e.
making $\Q$ less credible; see Fig.~\ref{fig_6_a_b}.

{\it (ii)} Decrease the width of the initial opinion of $\P$.

{\it (iii)} Decrease $\epsilon$ making $\Q$ more credible. In this
last case, the second peak of $\p(x)$ (the one close to the anchor of
$\Q$) will be dominant; see Fig.~\ref{fig_6_a_c}.

To understand the mechanism of the cognitive dissonance as described
by this model, let us start from (\ref{cruz}) and assume for
simplicity that the opinion of $\Q$ is certain: $q_k=0$ for $k\not =l$
and $q_l=1$. We get from (\ref{3}):
\begin{eqnarray}
  \label{mekh}
&&  \p_{k}=\frac{p_{k}}{1-p_l+p_l\sqrt{\epsilon+(1-\epsilon)p_l^{-1}}}~~{\rm
    for}~~ k\not=l, \\
&&  \p_{l}=\frac{p_l\sqrt{\epsilon+(1-\epsilon)p_l^{-1}}}
  {1-p_l+p_l\sqrt{\epsilon+(1-\epsilon)p_l^{-1}}}.
\end{eqnarray}
Now $\p_l/p_l>1>\p_{k}/p_{k}$, where $k\not=l$; hence even if $l$ was
on the tail of $\{p_k\}_{k=1}^N$, it is possible to make it a local
(or even the global) maximum of $\{\p_k\}_{k=1}^N$ provided that
$\epsilon$ is not close to $1$.}

The existence of at least two widely different probable opinions is
only one aspect of cognitive dissonance \cite{aronson,disso}. Another
aspect (sometimes called Freud-Festinger's law) is that people tend to
avoid cognitive dissonance: if in their action they choose one of the
two options (i.e. one of two peaks of the subjective probability),
they re-write the history of their opinion revision so that the chosen
option becomes the most probable one \cite{aronson,disso}. This aspect
of cognitive dissonance found applications in economics and decision
making \cite{akela,yariv}. The above points {\it (i)--(iii)} provide
concrete scenarios for a such re-writing.

\section*{\large Repeated persuasion}
\label{sec:sequential}

Here we analyze the opinion dynamics under repeated persuasion
attempts. Our motivation for studying this problem is that repeated
exposure to the same opinion is generally believed to be more
persuasive than a single exposure.
  
Under certain conditions ($p_kq_k\neq 0$, for all $k$ and
$1>\epsilon>0$) we show that the target opinion converges to the
persuading opinion after sufficient number of repetition. Below we
also examine how exactly this convergence takes place.

Assume that $\P$ revises his opinion repeatedly with the same opinion
of $\Q$.  Eq.~(\ref{3}) implies ($1\leq k\leq N$)
\BEA
\label{33}
p^{[n+1]}_k = \frac{{\sqrt{p^{[n]}_k \, [\epsilon p^{[n]}_k+
    (1-\epsilon)q_k) ]}}}{\sum_{l=1}^N 
{\sqrt{p^{[n]}_l \, [\epsilon p^{[n]}_l+
    (1-\epsilon)q_l) ]}}  }, 
\EEA 
where $1>\epsilon>0$, and $n=1,2,...$ is the discrete time.
For simplicity, we assume
\BEA
\label{ludmo}
p^{[1]}_k\equiv p_k>0, ~~~ q_k>0~~{\rm for}~~ 1\leq k\leq N.  \EEA 

Eq.~(\ref{33}) admits only one fixed point $q= \{q_k\}_{k=1}^N$.
Section VI of Supporting Information shows that for any convex,
$\frac{\d^2 f(y)} {\d y^2}\geq 0$, function $f(y)$ one has
\begin{eqnarray}
\label{sanik}
&&  \Phi[p^{[n+1]};q]\leq \Phi[p^{[n]};q], \\
&&  \Phi[p;q]\equiv {\sum}_{k=1}^N q_k f(p_k/q_k).
  \label{lyap}
\end{eqnarray}
Hence $\Phi[p;q]$ is a Lyapunov function of (\ref{33}).  Since
$\Phi[p;q]$ is a convex function of $p$, $\Phi[p;q]\geq
f(1)=\Phi[q;q]$ and $f(1)$ is the unique global minimum of
$\Phi[p;q]$. Section VI of Supporting Information shows that the
equality sign in (\ref{lyap}) holds ony for $p^{[n+1]}=p^{[n]}$. Thus
$\Phi[p^{[n]};q]$ monotonically decays to $f(1)=\Phi[q;q]$ showing
that the fixed point $q$ is globally stable. More generally, the
convergence reads: $p^{[n]}_k\to \zeta[p^{[1]}_k] q_k/\sum_{l=1}^N
\zeta[p^{[1]}_l]\, q_l$, where $\zeta(x>0)=1$ and $\zeta(0)=0$.

To illustrate (\ref{sanik}, \ref{lyap}), one can take
$f(y)=-\sqrt{y}$. Then (\ref{sanik}) amounts to decaying Hellinger
distance (\ref{hellinger}).  Many other reasonable measures of
distance are obtained under various choices of $f$.  { For instance,
  $f(y)=|y-1|$ amounts to decaying total variation distance
  (\ref{totality}), while $f(y)=-\ln y$ leads to the decaying relative
  entropy (Kullback-Leibler entropy).}

As expected, $0<\epsilon<1$ influences the convergence time. We
checked that this time is an increasing function of $\epsilon$, as
expected. In section VI of Supporting Information we also show
that the convergence to the fixed point respects the Le Chatelier
principle known in thermodynamics \cite{chat}: the probabilities of
those events that are overestimated from the viewpoint of $\Q$
(i.e. $p^{[1]}_k>q_k$) tend to decay in the discrete time. Likewise,
probabilities of the underestimated events (i.e. $p^{[1]}_k<q_k$)
increase in time.

Let us consider the Hellinger distance $h_n=h[p^{[n+1]},p^{[n]}]$
between two consecutive opinions of $\P$ evolving as in (\ref{33}). It
is now possible that
\begin{eqnarray}
  \label{gorgona}
{\rm max}_{1\leq n<\infty} [h_n]=h_m\not =h_1,
\end{eqnarray}
i.e. the largest change of the opinion of $\P$ comes not from the
first, but from one of intermediate persuasions. A simple example of
this situation is realized for $N=3$, an initial probability vector
$p=(0.98,0.01,0.01)$ and $q=(0.01,0.01,0.98)$ in (\ref{ludmo}). We
then apply (\ref{33}) under $\epsilon=0.5$. The consecutive Hellinger
distances read $h_1=0.1456<h_2=0.1567>h_3=0.1295>h_4...$.  Hence the
second persuasion changes the opinion more than others. For this to
hold, the initial opinion $p$ of $\P$ has to be far from the opinion
$q$ of $\Q$. Otherwise, we get a more expected behavior
$h_1>h_2>h_3>h_4...$ meaning that the first persuasion leads to the
largest change.

(The message of (\ref{gorgona}) is confirmed by using the discrete version
$\delta[p,q]=\frac{1}{2}\sum_k|p_k-q_k|$ of the distance
(\ref{totality}). Define $\delta_n=\delta[p^{[n+1]},p^{[n]}]$. Then
with $p=(0.98,0.01,0.01)$ and $q=(0.01,0.01,0.98)$ we get
$\delta_1=0.0834$, $\delta_2=0.1636$, $\delta_3=0.1717$,
$\delta_4=0.1444$.)

We conclude by stressing that while repeated persuasions drive the
opinion to its fixed point monotonically in the number of repetitions,
it is generally not true that the first persuasion causes the largest
opinion change, i.e. the law of diminishing returns does not hold. To
obtain the largest opinion change, one should carefully choose the
number of repetitions. 

Finally, note that the framework of (\ref{33}) can be applied to
studying mutual persuasion (consensus reaching). This is described in
Section VII of Supporting Information; see also \cite{curtis} in
this context.

\section*{\large Boomerang (backfire) effect} 
\label{boom}

\subsection*{Definition of the effect}

The {\it boomerang} or {\it backfire} effect refers to the empirical
observation that sometimes persuasion yields the opposite effect: the
persuaded agent $\P$ moves his opinion away from the opinion of the
persuading agent, $\Q$, i.e. he enforces his old opinion
\cite{boomerang,boomerang_red,nyhan,Whittaker1963}. Early literature
on social psychology proposed that the boomerang effect may be due to
persuading opinions placed in the latitude of rejection
\cite{Whittaker1963}, but this was not confirmed experimentally
\cite{kaplo}.

{ Experimental studies indicate that the boomerang effect is
  frequently related with opinion formation in an affective state,
  where there are emotional reasons for (not) changing the
  opinion. For example, a clear evidence of the boomerang effect is
  observed when the persuasion contains insulting language
  \cite{abelson}.} Another interesting example is when the subjects
had already announced their opinion publicly, and were not only
reluctant to change it (as for the usual conservatism), but even
enforced it on the light of the contrary evidence \cite{boomerang} (in
these experiments, the subjects who did not make their opinion public
behaved without the boomerang effect). A similar situation is realized
for voters who decided to support a certain candidate. After hearing
that the candidate is criticized, the voters display a boomerang
response to this criticism and thereby increase their support
\cite{nyhan,boomerang_red}.

\subsection*{Opinion revision rule}

We now suggest a simple modification of our model  that  accounts for the
basic phenomenology of the boomerang effect.

 { Recall our discussion (around (\ref{axiom6.1})) of various
  psychological and social factors that can contribute into the weight
  $\epsilon$. In particular, increasing the credibility of $\Q$ leads
  to a larger $1-\epsilon$. Imagine now that $\Q$ has such a low
  credibility that}
\begin{eqnarray}
  \label{bo}
  \epsilon>1.
\end{eqnarray}
Recall that $\epsilon=1$ means a special point, where no change of
opinion of $\P$ is possible whatsoever; cf. (\ref{3}).

After analytical continuation of (\ref{3}) for $\epsilon > 1$, the
opinion revision rule reads
\begin{eqnarray}
\label{b3}
\p_k=\frac{\sqrt{p_k |\epsilon p_k+ (1-\epsilon)q_k |}}{{\sum}_{l=1}^N 
\sqrt{p_l |\epsilon p_l+ (1-\epsilon)q_l |}},
\end{eqnarray}
with obvious generalization to probability densities. The absolute
values in (\ref{b3}) are necessary to ensure the positivity of
probabilities. 

{ It is possible to {\it derive} (rather simply postulate)
  (\ref{b3}). Toward this end, let us return to the point {\bf 6.1}
  and (\ref{axiom6.1}). During the opinion combination step, $\P$
  forms $\epsilon p_k+(1-\epsilon)q_k$ which in view of $\epsilon>1$
  can take negative values and hence is a signed measure. Signed
  measures have all formal features of probability besides positivity
  \cite{bart,allen,hungar,burgin}; see section V of Supporting
  Information for details. There is no generally accepted
  probabilistic interpretation of signed measures, but in section V of
  Supporting Information we make a step towards such an interpretaion.
  There we propose to look at a signed measure as a partial
  expectation value defined via joint probability of the world's
  states and certain hidden degrees of freedom (e.g. emotional
  states). After plausible assumptions, the marginal probability of
  the world's states is deduced to be
\begin{eqnarray}
  \label{negus1}
\widehat{p}_k=|\epsilon p_k+(1-\epsilon)q_k|\left/
\sum_{l=1}^N|\epsilon p_k+(1-\epsilon)q_k| \right. ,
\end{eqnarray}
We obtain (\ref{b3})
after applying (\ref{axiom6.2}, \ref{upo}) to (\ref{negus1}). }

\subsection*{Scenarios of opinion change}

According to (\ref{bo}, \ref{b3}) those opinions of $\P$ which are
within the overlap between $p$ and $q$ (i.e $p_kq_k\not\approx 0$) get
their probability decreased if $p_k/q_k\approx
(\epsilon-1)/\epsilon<1$, i.e. if the initial $p_k$ was already
smaller than $q_k$. In this sense, $\P$ moves his opinion away from
that of $\Q$. Hence for continuous densities $p(x)$ and
  $q(x)$ there will be a point $x_0$, { where $\p(x_0)$ is
    close to $0$. This point is seen in Figs.~\ref{fig_4} and
    \ref{fig_3_b}}.

Fig.~\ref{fig_4} illustrates the shape of $\p(x)$ produced by
(\ref{b3}) for initially Gaussian opinions (\ref{38}) of $\P$ and
$\Q$. It is seen that $\P$'s anchor moves away from $\Q$'s anchor,
while the width of $\p(x)$ around the anchor is more narrow than that
of $p(x)$; cf. with Fig.~\ref{fig_3_a}. To illustrate
these points analytically, we return to (\ref{pushi}, \ref{lino},
\ref{linux}) that for $v_\P\approx v_\Q$ and $m_\P\approx m_\Q$
predict $m_{\tilde{\P}}-m_{\P}=\frac{1-\epsilon}{2} (m_{\Q}-m_{\P})$:
for $\epsilon>1$, $\P$'s anchor drifts away from $\Q$'s anchor.

Likewise, whenever the two anchors are equal, $m_\P = m_\Q$,
inequality (\ref{sakhalin}) is reversed in the boomerang regime
(\ref{bo}).

Let us now consider the impact of the presentation order under this
settings. We saw that for $0<\epsilon<1$ the model predicts recency
effect. For $1\lesssim \epsilon$ we expect the recency effect is
still effective as implied by the argument (\ref{gilgamesh}). However,
the situation changes drastically for $\epsilon$ sufficiently larger
than $1$, as indicated in Fig.~\ref{fig_3_b}. Now the primacy effect
dominates, i.e. instead of (\ref{order_presentation}) we get the
opposite inequality.  Fig.~\ref{fig_3_b} also shows that interaction
with two contradicting opinions (in the boomerang regime) enforces the
initial anchor of $\P$.

To understand the primacy-recency effect analytically, consider
the example (\ref{ord}), and recall that $\P$ interacts first with $\Q$
and then with $\Q'$ with the same parameter $\epsilon$. The resulting
opinion $ p(q,q')$ of $\P$ reads:
\begin{eqnarray}
  \label{khorezm1}
  p(q,q') = \left(\,\frac{g(\epsilon)}{g(\epsilon)+1},\,\,
  \frac{1}{g(\epsilon)+1}\,\right), \\
  g(\epsilon)=\sqrt{\,\left| \frac{\sqrt{\epsilon}
        +(1-\epsilon)\sqrt{|2-\epsilon|}   }{\sqrt{\epsilon}
          (2-\epsilon)}    
\right|    \,}.
  \label{khorezm2}
\end{eqnarray}
Fig.~\ref{fig_5} shows how
$p_1(q,q')=\frac{g(\epsilon)}{g(\epsilon)+1} $ behaves as a function
of $\epsilon$. The recency effect holds for $\epsilon< 2+\sqrt{2}$;
for $\epsilon> 2+\sqrt{2}$ we get primacy. Similar results are
obtained for initially Gaussian opinions. 

Thus, in the present model, the primacy effect (relevance of the first
opinion) can be related to the boomerang effect.

We now examine the emergence of cognitive dissonance in the boomerang
regime $\epsilon>1$. Our results indicate that in this regime the
agent is more susceptible to cognitive dissonance;
cf. Fig.~\ref{fig_4} with Figs.~\ref{fig_1}. The mechanism of the
increased susceptibility is explained in Fig.~\ref{fig_4}: $\P$'s
opinion splits easier, since the probability mass moves away (in
different directions) from the anchor of $\Q$.

\comment{
Furthermore, a stronger form of cognitive dissonance is possible, where the
anchor moves in one direction (away from $\Q$), while the probability
density moves towards $\Q$ in terms of distance. This is illustrated by
Fig.~\ref{fig_7}, where for $\epsilon>6$, $\p(x)$ (opinion of $\P$
after one interaction) moves closer to $q(x)$ (as measured by the
Hellinger distance), while the anchor of $\P$ moves away from that of
$\Q$.}

Let us now assume that $\P$ repeatedly interacts  with the same opinion of
$\Q$ [cf. (\ref{33})]:
\begin{eqnarray}
\label{gonzal}
p^{[n+1]}(x)=\frac{{\sqrt{p^{[n]}(x) \, |\epsilon p^{[n]}(x)+
      (1-\epsilon)q(x) |}} }{ \int\d x'{\sqrt{p^{[n]}(x') \, |\epsilon
      p^{[n]}(x')+ (1-\epsilon)q(x') |}}}, 
\end{eqnarray}
where $n=1,2,...$ is the discrete time. Starting from initially
Gaussian opinion, $\P$ develops two well-separated peaks, which is
another manifestation of cognitive dissonance: the smaller peak moves
towards the anchor of $\Q$ and finally places itself within the
acceptance latitude of $\Q$, where the larger peak becomes more narrow
and drifts away from $q(x)$; see Fig.~\ref{fig_6_b}. After many
iterations ($\simeq 10^3$ for parameters of Fig.~\ref{fig_6_b}) the
larger peak places itself within the rejection latitude of $\Q$, at
which point $p^{[n]}(x)$ stops changing (stationary opinion). The
above scenario suggests that in the boomerang regime there is a finite
probability that the target agent will eventually be persuaded after
repeated exposure to the same opinion.

{ Let us mention an experimental work that is relevant to our
  discussion above.  Ref.~\cite{boomerang_red} carried out experiments
  with subjects displaying boomerang effect, where each subject was
  exposed to sufficiently many different (but still similar)
  persuasive opinions. It was found that, sooner or later, the
  subjects exit the boomerang regime, i.e. they start to follow the
  persuasion \cite{boomerang_red}. Our set-up is somewhat different in
  that the subject ($\P$) is repeatedly exposed to the same persuading
  opinion. Modulo this difference, our conclusion is similar to the
  experimental finding: the agent starts following the persuasion with
  a certain probability. }

\section*{\large Discussion}
\label{verch}

We presented a new model for opinion revision in the presence of
confirmation bias. The model has three inputs: the subjective
probabilistic opinions of the target agent $\P$ and a persuading
(advising) agent $\Q$, and the weight of $\Q$ as perceived by
$\P$. 

{The basic idea of the opinion revision rule is that no opinion change
  is expected if the persuasion is either too far or too close to the
  already existing opinion \cite{shreider,beim,ontology}.}  The
opinion revision rule is not Bayesian, because the standard Bayesian
approach does not apply to processes of persuasion and advising; see
the second section for more details.

The model accounts for several key empirical observations reported in
social psychology and quantitatively interpreted within the social
judgment theory. In particular, the model allows to formalize the
concept of opinion latitudes, explains the structure of the weighted
average approach to opinion formation, and relates the initial
discrepancy (between the opinions of $\P$ and $\Q$) to the magnitude
of the opinion change (shown by $\P$). In all these cases our model
extends and clarifies previous empiric results, e.g. it elucidates the
difference between monotonic and non-monotonic change-discrepancy
relations, identifies conditions under which the opinion change is
sudden, as well as provides a deeper perspective on the weighted
average approach.

New effects predicted by the model are summarized as follows. 

{\it (i)} For the order of presentation set-up (and outside of the
boomerang regime) the model displays recency effect. We suggested that
the standard argument that relates confirmation bias to the primacy
effect does not work in this model.  In this context we recall a
widespread viewpoint that {\it both} recency and primacy relate to
(normative) irrationality; see e.g. \cite{baron}. However, the
information which came later is generally more relevant for predicting
future. Hence recency can be more rational than primacy.

In many experimental set-ups the recency changes to primacy upon
increasing the retention time; see e.g. \cite{wright}. Our model
demonstrates the primacy effect only in the boomerang regime
(i.e. only in the special case). Hence, in future it needs to be
extended by involving additional mechanisms, e.g. those related to
``long-term memory'' processes which could be responsible for the
above experimental fact. Recall in this context there are several
other theoretical approaches that address the primacy-recency
difference \cite{Hogarth1992,atman,khren,pothos,buse}.

{\it (ii)} The model can be used to describe the phenomenon of
cognitive dissonance and to formalize the main scenario of its
emergence.

{\it (iii)} Repeated persuasions display several
features implying monotonous change of the target opinion towards the
persuading opinion. However, the opinion changes do not obey the law of
diminishing returns, or in other words, the first persuasion is not
always leads to the largest change. These findings may contribute to
better understanding the widespread use of repeated persuasions.

{\it (iv)} We proposed that the boomerang effect is related to the
limit of this model, where the credibility of persuasion is (very)
low. A straightforward implementation of this assumption led us to a
revision rule that does describe several key observational features of
the boomerang effect and predicts new ones; e.g. that in the boomerang
regime the agent can be prone to primacy effect and to cognitive
dissonance. There are, however, several open problems with the opinion
revision rule in the boomerang regime. They should motivate future
developments of this model. One problem concerns relations of the
revision rule with signed measures that at a preliminary level were
outlined in section V of Supporting Information. Another problem
is that the revision rule in the boomerang regime (and only there) is
not completely smooth, { since it includes the function
  $|x|$, whose second derivative is singular}. We do hope to clarify
these points in future.

In this paper we restricted ourselves by studying few (two or three)
interacting agents with opinions described via subjective
probabilities. However, these probabilities can also represent an
ensemble of agents each one having a fixed (single) opinion, a useful
viewpoint on subjective probabilities advocated in
Ref.~\cite{jaynes}. In future we plan to explore this point and also
address the opinion dynamics for collectives of agents. This last
aspect was recently extensively studied via methods of statistical
physics; see \cite{fortunato,stauffer} for reviews.

\section*{Acknowledgments}

We thank Seth Frey for useful remarks and suggestions. 



\begin{figure*}[htbp]
\section*{\large Figures}
\centering
\subfigure[]{
     \label{fig_1_a}
    \includegraphics[width=0.55\columnwidth]{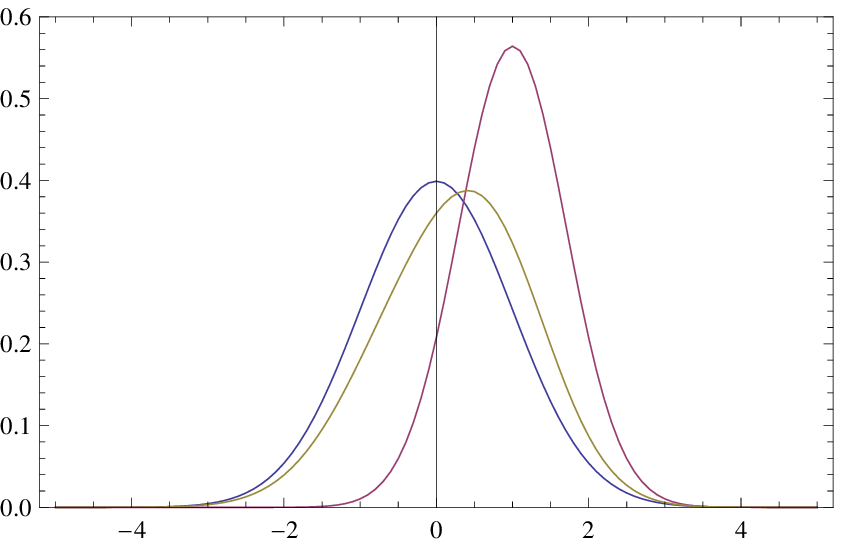} 
    } 
    \subfigure[]{
     \label{fig_1_b}
    \includegraphics[width=0.55\columnwidth]{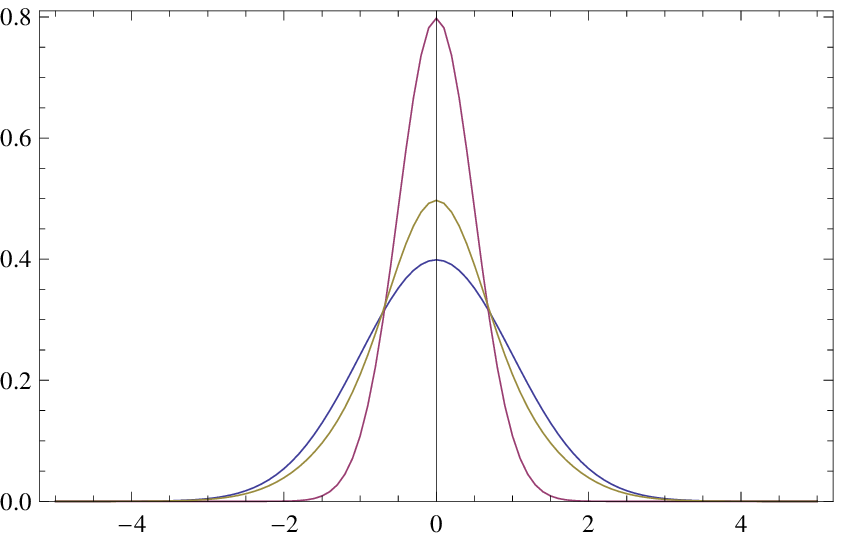} 
    }
    \subfigure[]{
     \label{fig_1_c}
   \includegraphics[width=0.55\columnwidth]{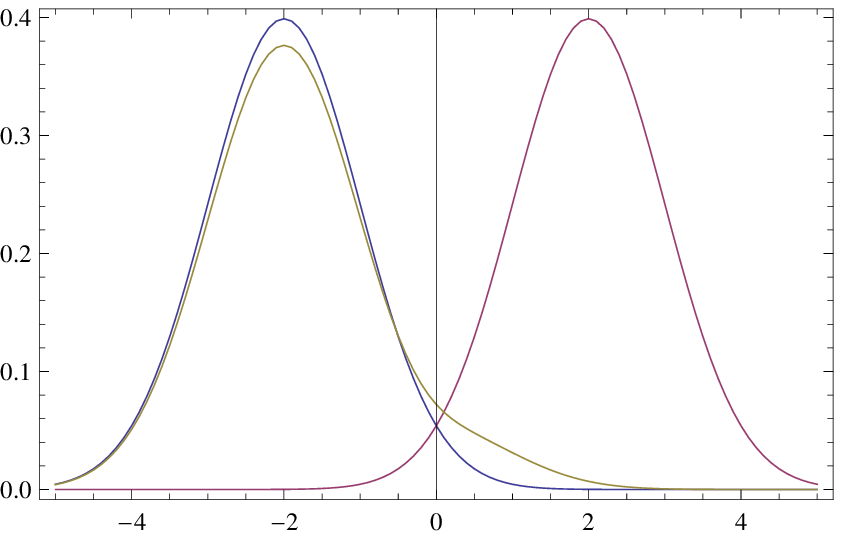} 
    }
    \caption{{\bf Opinions described via Gaussian densities} (\ref{38}).\\
      The initial opinion of $\P$ is described by Gaussian probability
      density $p(x)$ (blue curve) centered at zero; see
      (\ref{38}). The opinion of $\Q$ amounts to Gaussian probability
      density $q(x)$ (purple curve) centered at a positive value.  For
      all three figures continuous density $f(x)$ ($f=p,\, q,\, \p$)
      were approximated by $100$ points $\{f(x_k)\}_{k=1}^{100}$,
      $x_{k+1}-x_k=0.1$. The resulting opinion $\p(x)$ of $\P$ is
      given by (\ref{30}) with $\epsilon=0.5$ (olive curve).
      \\
      (a) The opinion of $\P$ moves towards that of $\Q$;
      $m_\P=0$, $\sigma_\P=1$, $m_\Q=1$, $\sigma_\Q=0.5$.\\
      (b) The maximally probable opinion of $\P$ is reinforced;
      $m_\P=0$, $\sigma_\P=1$, $m_\Q=0$, $\sigma_\Q=0.25$.\\
      { (c) The change of the opinion of $\P$ is relatively
        small provided that the Gaussian densities overlap only in the
        region of non-commitment; cf. (\ref{lat}), (\ref{lat_r}).\\
        Whenever the densities overlap only within the rejection range
        the difference between $p(x)$ and $\p(x)$ is not visible by
        eyes. For example, if $p(x)$ and $q(x)$ are Gaussian with,
        respectively, $m_\P=-3$, $m_\Q=3$, $v_\P=v_\Q=1$, the
        Hellinger distance (see (\ref{hellinger}) for definition)
        $h[p,q]=0.99$ is close to maximally far, while the opinion
        change is small: $h[p,\p]=3.48\times 10^{-2}$.  }}
\label{fig_1}
\end{figure*}

\begin{figure*}[htbp]
  \includegraphics[width=6cm]{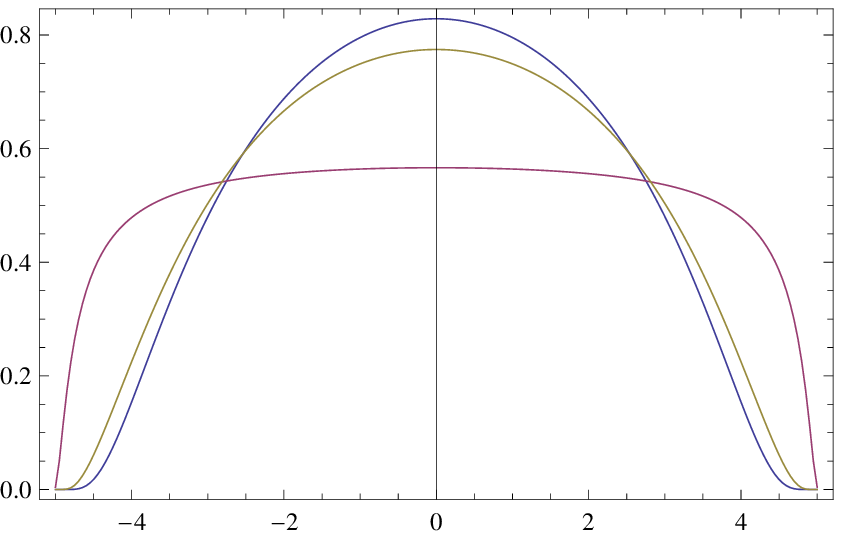} 
  \caption{{{\bf Opinions described via bump densities}
    (\ref{bump}).\\ Blue curve: the initial opinion of $\P$ given by
    (\ref{bump}) with $b=1$. Purple curve: the opinion of $\Q$
    described by (\ref{bump}) with $b=0.001$. Olive curve: the
    resulting opinion of $\P$ obtained via (\ref{30}) with
    $\epsilon=0.5$. } }
\label{fig_1_1}
\end{figure*}

\begin{figure*}[htbp]
\centering
\subfigure[]{
     \label{fig_2_a}
     \includegraphics[width=5.2cm]{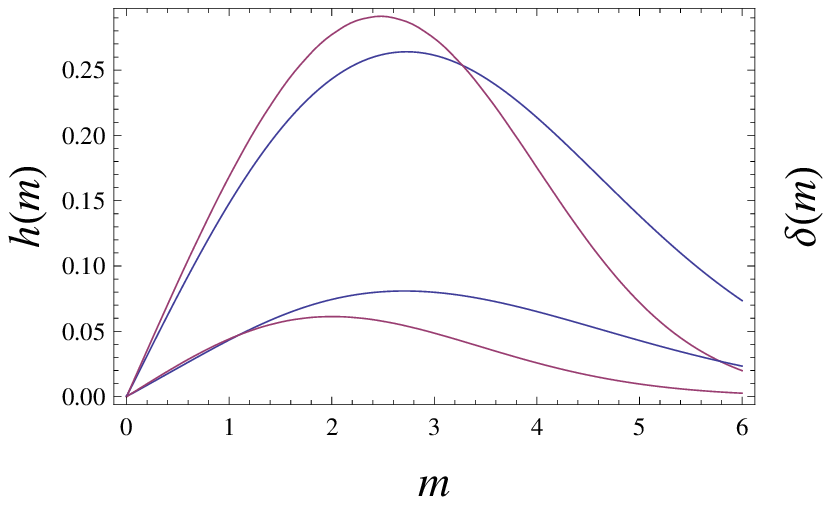}
        } 
    \subfigure[]{
     \label{fig_2_b}
    \includegraphics[width=5.2cm]{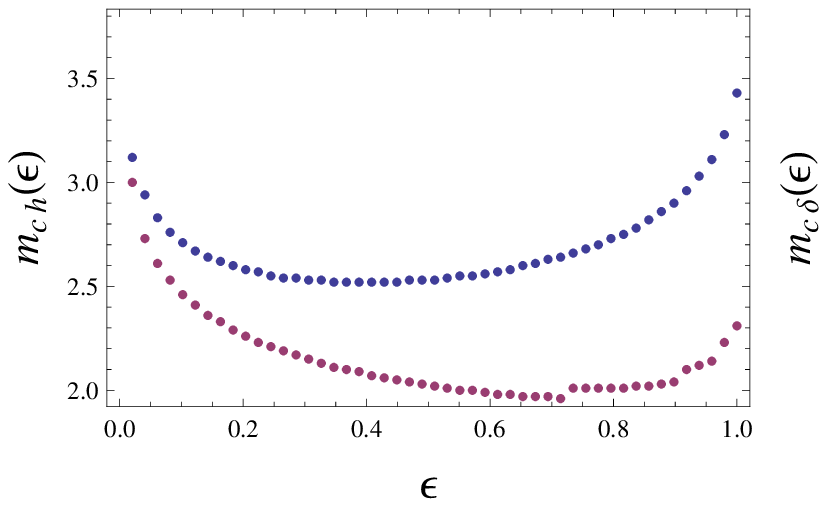}    
    }
\subfigure[]{
     \label{fig_2_c}
     \includegraphics[width=4.7cm]
{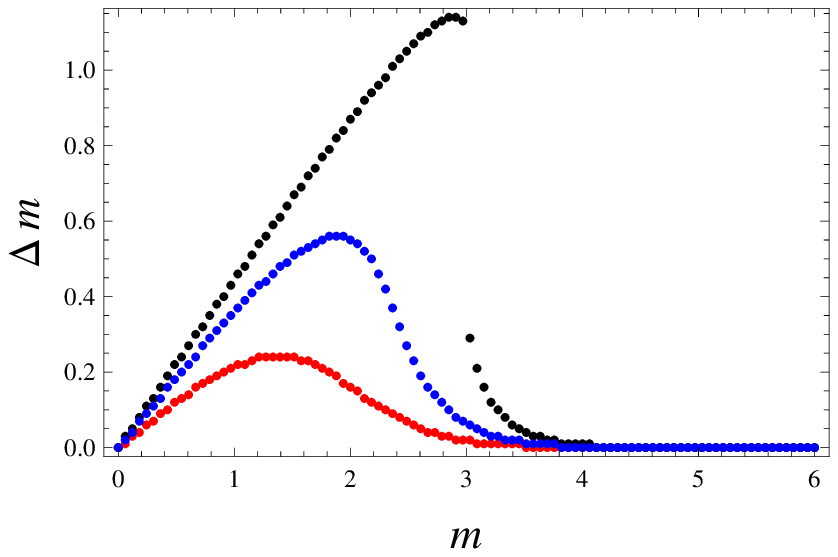}
        } 
        \caption{{\bf Opinion change versus discrepancy.} \\
          (a) The opinion change is quantified via the Hellinger
          distance $h=h[p,\p]$ between the old and new opinion of $\P$
          (blue curves); see (\ref{hellinger}) for the definition. For
          comparison we also include the total variance distance
          $\delta=\delta[p,\p]$ (purple curves); see
          (\ref{totality}). These two distances are plotted versus the
          discrepancy $m=|m_\P-m_\Q|$.  The initial opinion of the
          agent $\P$ is Gaussian with $m_\P=0$ and $v_\P=1$; see
          (\ref{38}). The opinion of $\Q$ is Gaussian with $m_\Q=m$
          and $v_\Q=1$. Thus $m$ quantifies the initial distance
          between the opinions of $\P$ and $\Q$. The final opinion
          $\p(x)$ is given by (\ref{3}). Different curves correspond
          to different $\epsilon$.\\ Blue curves: $h(m)=h[p,\p]$ for
          $\epsilon=0.1$ (upper curve) and $\epsilon=0.5$ (lower
          curve). Purple curves: $\delta(m)=\delta[p,\p]$ for
          $\epsilon=0.1$ (upper curve) and $\epsilon=0.5$ (lower
          curve).  The maximum of $h(m)$ ($\delta(m)$) is
          reached at $m_{c\,h}$ ($m_{c\,\delta}$).      \\
           { (b) $m_{c\,h}$ ($m_{c\,\delta}$) is the point
            where $h(m)$ ($\delta(m)$) achieves its maximum as a
            function of $m$.  Blues points: $m_{c\, h}(\epsilon)$
            versus $\epsilon$ for same parameters as in (a). $m_{c\,
              h}(\epsilon)$ grows both for $\epsilon\to 1$ and
            $\epsilon\to 0$, e.g. $m_{c\, h}(0.01)=3.29972$, $m_{c\,
              h}(0.0001)=4.53052$, $m_{c\, h}(0.9)=2.94933$, $m_{c\,
              h}(0.999)=4.12861$.  Purple points: $m_{c\,
              \delta}(\epsilon)$ versus $\epsilon$
            for same parameters as in (a).  \\
            (c) The difference of the anchors (maximally probable
            values) $\Delta m= m_{\widetilde{\P}}-m_\P$ versus
            $m_\Q=m$ for the initial opinions of $\P$ and $\Q$ given
            by (\ref{38}) under $m_\P=0$, $v_\P=1$, $m_\Q=m$ and
            $v_\Q=1$. The final opinion $\p(x)$ of $\P$ (and its
            maximally probable value $m_{\widetilde{\P}}$) if found
            from (\ref{3}) under $\epsilon=0.1$ (black points),
            $\epsilon=0.25$ (blue points) and $\epsilon=0.5$ (red
            points).  }}
\label{fig_2}
\end{figure*}

\begin{figure*}[htbp]
  \includegraphics[width=6cm]{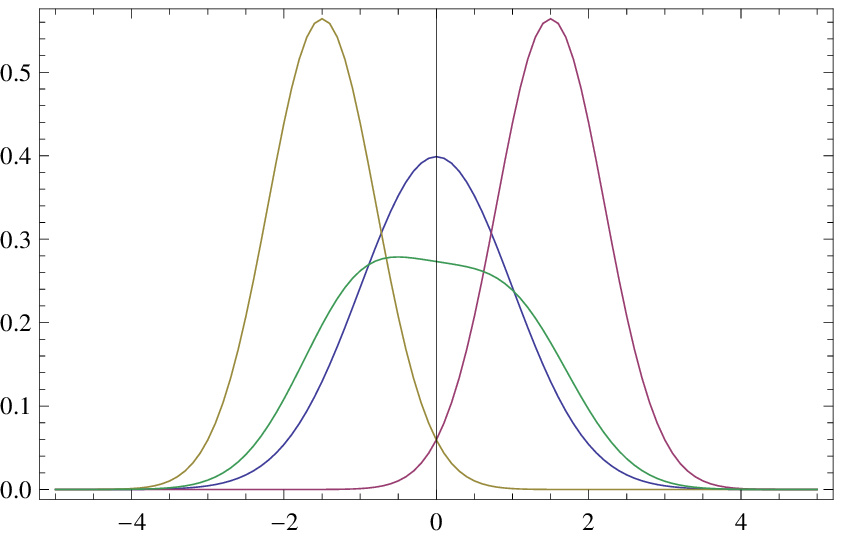}
    \caption{{\bf Order of presentation effect.}\\
      Blue curve: The initial opinion of $\P$ is described by Gaussian
      probability density $p(x)$ with $m_\P=0$ and $v_\P=1$; see
      (\ref{38}). Purple (resp. olive) curve: the initial opinion of
      $\Q$ (resp. $\Q'$) are given by (\ref{38}) with $m_\Q=1.5$
      (resp. $m_{\Q'}=-1.5$) and $v_\Q=0.5$
      (resp. $v_{\Q'}=0.5$). Green curve: the resulting opinion of
      $\P$ after interacting first with $\Q$ and then with $\Q'$. Both
      interactions use $\epsilon=0.5$. The final opinion of $\P$ is
      inclined to the most recent opinion (that of $\Q'$) both with
      respect to its maximally probable value and distance. The final
      opinion of $\P$ has a larger width than the initial one.  }
\label{fig_3_a}
\end{figure*}

\begin{figure*}[htbp]
\centering
\subfigure[]{
     \label{fig_6_a_a}
     \includegraphics[width=5.5cm]{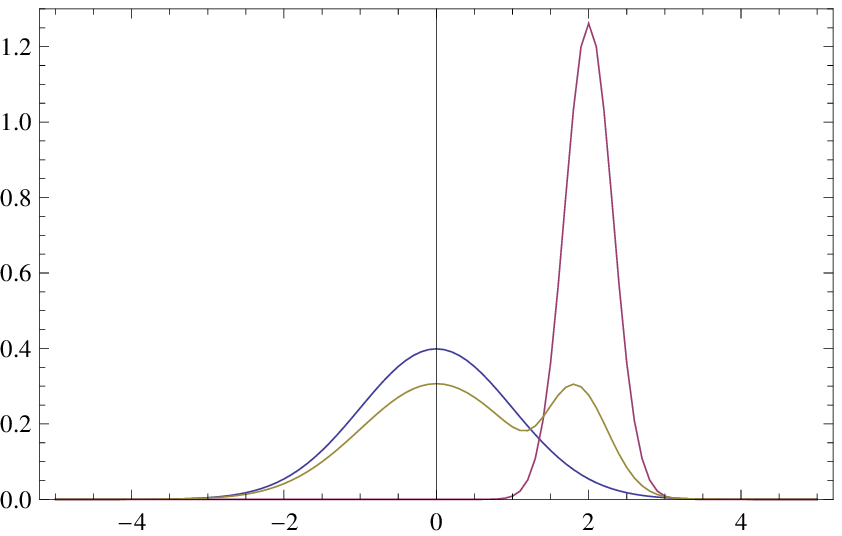} 
        } 
\subfigure[]{
     \label{fig_6_a_b}
     \includegraphics[width=5.5cm]{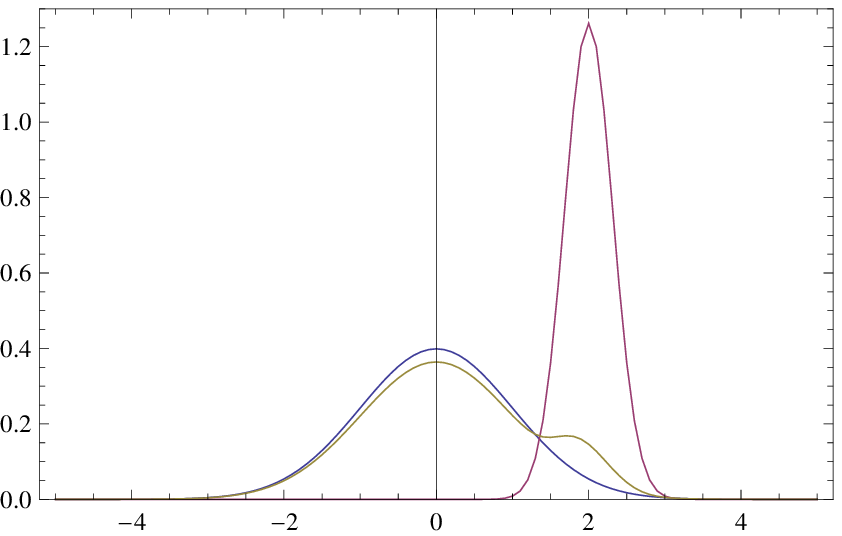} 
        } 
\subfigure[]{
     \label{fig_6_a_c}
     \includegraphics[width=5.5cm]{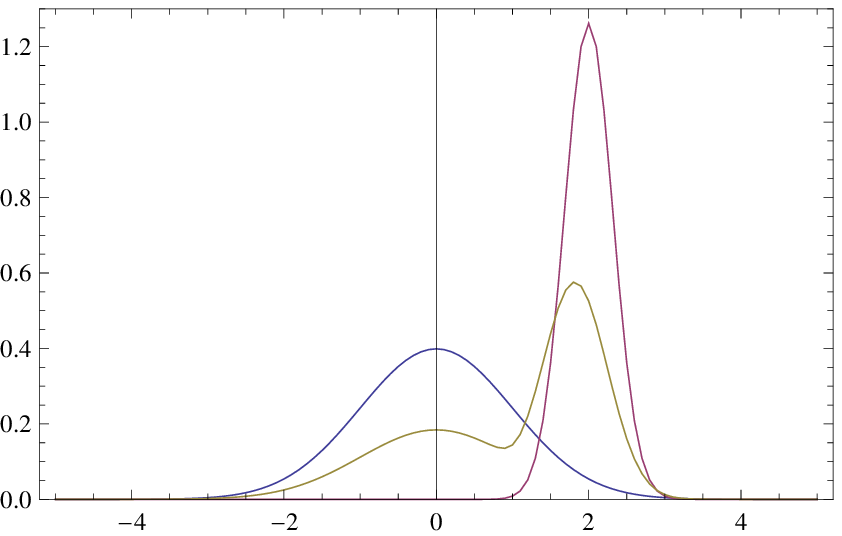} 
        } 
        \caption{{\bf Cognitive dissonance.}\\
          (a) Blue (resp. purple) curve: the initial opinion of agent
          $\P$ (resp. $\Q$) described by probability density $p(x)$
          (resp. $q(x)$). Olive curve: the final opinion $\p(x)$ of
          $\P$ as given by (\ref{30}) with $\epsilon=0.35$. Here
          $p(x)$ and $q(x)$ are defined by (\ref{38}) with $m_\P=0$,
          $v_\P=1$, $m_\Q=2$, $v_\Q=0.1$. The final opinion develops
          two peaks of comparable height (cognitive
          dissonance).\\
          (b) { Avoiding the cognitive dissonance due to a
            larger $\epsilon=0.75$: the second peak is much smaller
            (other parameters are those of (a)).\\
            (c) Avoiding the cognitive dissonance due to a smaller
            $\epsilon=0.05$: the first peak is much smaller (other
            parameters are those of (a)).}  }
     \label{fig_6_a}
\end{figure*}

\begin{figure*}[htbp]
     \includegraphics[width=6.6cm]{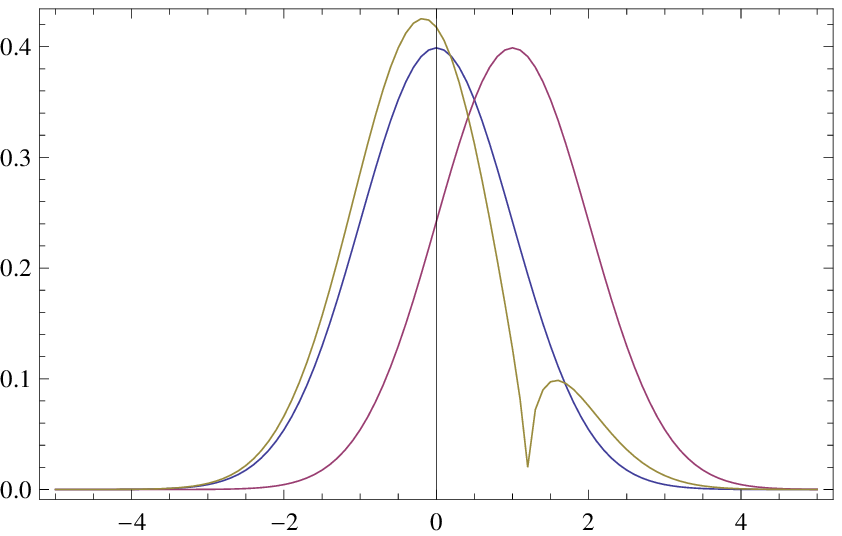}
     \caption{{\bf Opinion change in the boomerang regime.} \\
       Blue (resp. purple) curve: the initial opinion of agent $\P$
       (resp. $\Q$) described by probability density $p(x)$
       (resp. $q(x)$). Olive curve: the final opinion $\p(x)$ of $\P$
       given by (\ref{30}) with $\epsilon=2$. Here $p(x)$ and $q(x)$
       are given by (\ref{38}) with $m_\P=0$ and
       $v_\P=m_\Q=v_\Q=1$. The anchor (maximally probable opinion) of
       $\P$ not only moves away from the anchor of $\Q$; but it is
       also enhanced: the (biggest) peak of $\p(x)$ is larger than
       that of $p(x)$. The second (smaller) peak of $\p(x)$ arises
       because the initial probability of $\P$ located to the right
       from the anchor $m_\Q$ of $\Q$, moves away from $m_\Q$; $\p(x)$
       gets a local minimum close to $m_\Q$. }
\label{fig_4}
\end{figure*}

\begin{figure*}[htbp]
    \includegraphics[width=6cm]{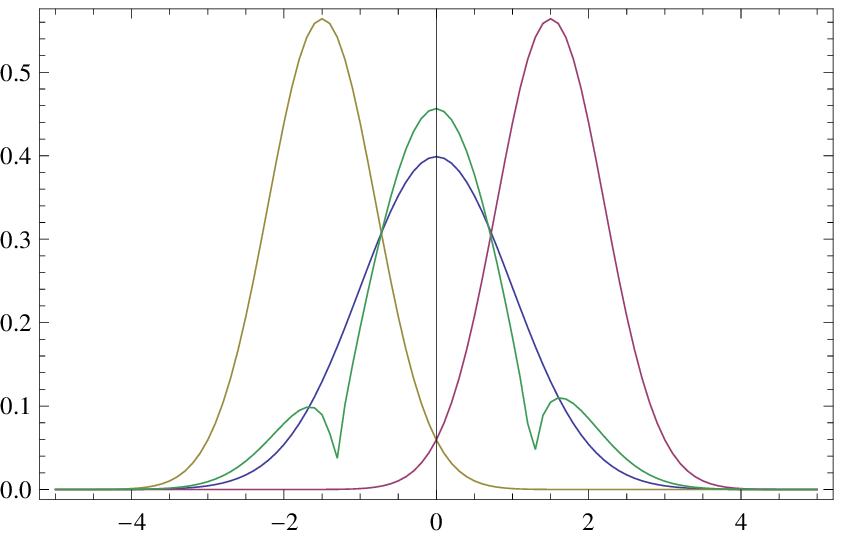}
    \caption{{\bf Order of presentation effect in the boomerang regime}.\\
      The same as in Fig.~\ref{fig_3_a} but for $\epsilon=1.5$
      (boomerang regime). Now the final opinion of $\P$ is inclined to
      the first opinion (that of $\Q$) with respect to the
      distance. The initial maximally probable opinion of $\P$ is
      still maximally probable. Moreover, its probability has
      increased and the width around it has decreased. The final
      opinion has 3 peaks. }
     \label{fig_3_b}
\end{figure*}

\begin{figure*}[htbp]
     \includegraphics[width=6.6cm]{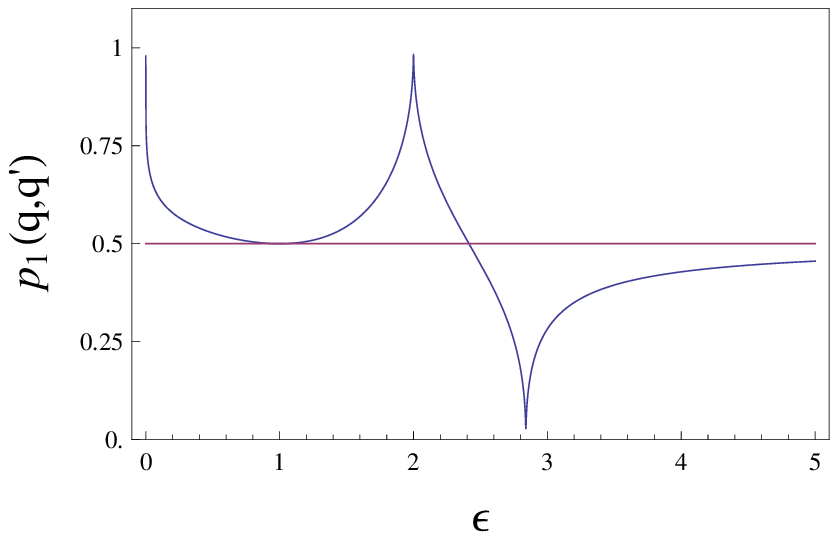}
     \caption{ {\bf Illustration of the order of presentation effect
         in the boomerang regime}.\\
       $p_1(q,q')=\frac{g(\epsilon)}{g(\epsilon)+1} $ given by
       (\ref{khorezm1}, \ref{khorezm2}) versus $\epsilon$.  }
\label{fig_5}
\end{figure*}

\begin{figure*}[htbp]
    \includegraphics[width=6.6cm]{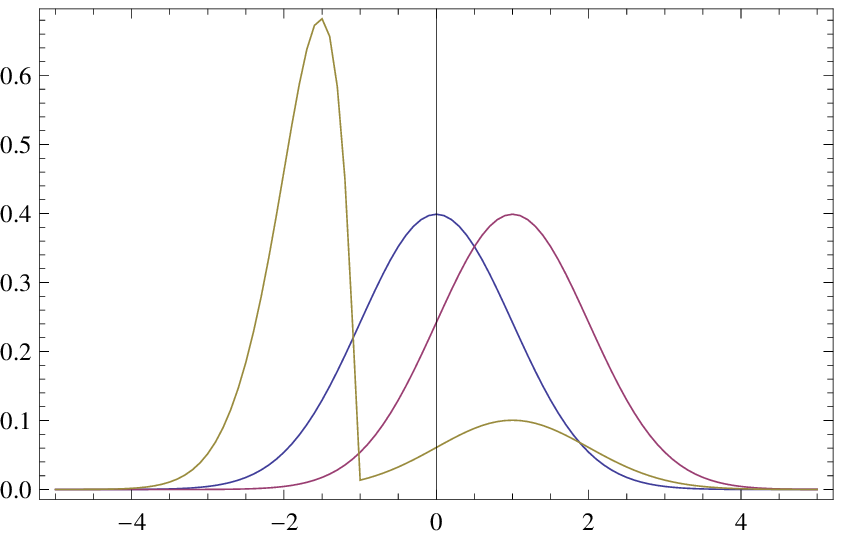} 
    \caption{
      {\bf Repeated persuasion in the boomerang regime.} \\
      Blue (resp. purple) curve: the initial opinion of agent $\P$
      (resp. $\Q$) described by probability density $p(x)$
      (resp. $q(x)$) as given by (\ref{38}) with $m_\P=0$, $v_\P=v_Q=1$,
      $m_\Q=1$. Olive curve: the opinion of $\P$ after 50 iterations
      (\ref{gonzal}) with $\epsilon=2$.  }
\label{fig_6_b}
\end{figure*}

\comment{
\begin{figure*}[htbp]
    \includegraphics[width=6.6cm]{3_boomerang_confirmation_bias.eps}
    \caption{ {\bf Opinion change after one interaction} (including the
      boomerang regime); see (\ref{3}, \ref{30}, \ref{b3}). The
      Hellinger distance difference $h[p,q]-h[\p,q]$ versus
      $\epsilon$. The initial opinions $p(x)$ and $q(x)$ are given by
      (\ref{38}) with $m_\P=0$, $v_\P=m_\Q=v_\Q=1$ (lower curve) and
      $m_\P=0$, $m_\Q=1.5$, $v_\P=v_\Q=1$ (upper
      curve). $h[p,q]-h[\p,q]>0$ means that the final opinion of $\P$
      is closer to that of $\Q$ in terms of the Hellinger distance.  }
\label{fig_7}
\end{figure*}
}


\clearpage

\begin{thebibliography}{99}

  \comment{: van Wilgenburg E, Elgar MA (2013) Confirmation Bias in
    Studies of Nestmate Recognition: A Cautionary Note for Research
    into the Behaviour of Animals. PLoS ONE 8(1):
    e53548. doi:10.1371/journal.pone.0053548

    Editor: Deborah M. Gordon, Stanford University

    Confirmation bias is a tendency of people to interpret information
    in a way that confirms their expectations. A long recognized
    phenomenon in human psychology

    Fanelli D (2010) Do Pressures to Publish Increase Scientists'
    Bias? An Empirical Support from US States Data. PLoS ONE 5(4):
    e10271. doi:10.1371/journal.pone.0010271

    Editor: Enrico Scalas
}


{
\bibitem{abelson}
Abelson RP (1986). Beliefs are like possessions. Journal for the
Theory of Social Behaviour 16: 223-250.}



\bibitem{akela}
Akerlof G, Dickens WT (1982). The economic consequences of cognitive
dissonance. Amer. Econ. Rev. 72: 307-319.

{
\bibitem{alc}Alchourr\'on CE, G\"ardenfors G, Makinsin D (1985). On the
  logic of theory change. J. Symb. Logic 50: 510-530.}


\bibitem{chat} Allahverdyan AE, Galstyan A (2011). Le
  Chatelier principle in replicator dynamics.  Physical
  Review E 84: 041117 

\bibitem[Allakhverdov and Gershkovich(2010)]{allakh}
Allakhverdov VM, Gershkovich VA (2010).
 Does consciousness exist? in what sense?
Integrative Psychological and Behavioral Science 44:
  340-347.


{
\bibitem{allen} Allen EH (1976). Negative Probabilities and the Uses
  of Signed Probability Theory. Philosophy of Science 43: 53-70.}


\bibitem{anderson} Anderson NN (1965). Primacy effect in
  personality impression formation using generalized order effect
  paradigm.  Journal of Personality and Social Psychology 2: 1-9. 

\bibitem{anderson_review}  { Anderson NH
    (1971). Integration theory and attitude change. Psychological
    Review 78: 171-206.}



\bibitem{anderson_book} Anderson NN (1981). Foundations of information
  integration theory. Academic Press, New York.

\bibitem[Aronson, 2007]{aronson} 
 Aronson E (2007). The Social Animal. Palgrave Macmillan, 10th
  revised edition.

\bibitem{atman}
Atmanspacher H, Roemer H (2012). Order effects in sequential
    measurements of non-commuting psychological observables. Journal
    of Mathematical Psychology 56: 274–280.

\bibitem{grif}
Austerweil JL, Griffiths TL (2011). Seeking
  confirmation is rational for deterministic hypotheses.  Cognitive
  Science 35: 499-526

\bibitem{baron} Baron J (2008). Thinking and deciding. Cambridge
  University Press, Cambridge.

{
\bibitem{bart} Bartlett MS (1945). Negative probability. Mathematical
  Proceedings of the Cambridge Philosophical Society 41: 71-73.}


\bibitem[beim Graben(2006)]{beim} beim Graben P (2006).  Pragmatic
  information in dynamic semantics.  Mind and Matter, 4: 169-193.

{
\bibitem{bernardo}Bernardo JM (1979). Expected information as
  expected utility. Annals of Statistics 7: 686-690.}


\bibitem[Bochner and Insko(1966)]{Bochner1966}
Bochner S, Insko CA (1966).
Communicator discrepancy, source credibility, and opinion change.
Journal of Personality and Social Psychology, 4: 133-140.

{
\bibitem{advising_review} Bonaccio S, Dalal RS (2006). Advice taking
  and decision-making: An integrative literature review, and
  implications for the organizational sciences. Organizational
  Behavior and Human Decision Processes 101: 127–151.}


{
\bibitem{burgin} Burgin M, Meissner G (2012). Negative probabilities
  in financial modeling. Wilmott Magazine 58: 60–65.}




\bibitem{fortunato}
  Castellano C, Fortunato S, Loreto V (2009). Statistical physics of
  social dynamics. Reviews of Modern Physics 81: 591-646.

\bibitem[Clemen and Winkler(1999)]{clemen_winkler} Clemen RT, Winkler
  RL (1999).  Combining probability distributions from experts in risk
  analysis.  Risk Analysis 19: 187-203.


{
\bibitem{cox}Cox RT (1946). Probability, Frequency and Reasonable
  Expectation. American Journal of Physics 14: 1-13.}


{
\bibitem{curtis}Curtis JP, Smith FT (2008).  Mathematical Models of
  Persuasion.  American Conference on Applied Mathematics (MATH '08),
  Harvard, Massachusetts: 60-65.}





\comment{
\bibitem[Cooke(1991)]{coo}
R.~Cooke.
\newblock \emph{Experts in uncertainty: Opinion and subjective probability in
  science}.
\newblock Oxford University Press, Oxford, UK, 1991.

\bibitem{topoi}R.M. Cooke, Topoi, {\bf 5}, 21 (1986). 
}

\bibitem[Darley and Gross(1983)]{darley_gross}
Darley JM, Gross PH (1983).
A hypothesis-confirming bias in labeling effects.
Journal of Personality and Social Psychology 44: 20-33.

\bibitem[Diaconis and Zabell(1982)]{diaconis_zabell}
Diaconis P, Zabell SL (1982).
Updating subjective probability.
Journal of the American Statistical Association, 77: 822-830.

\bibitem[Festinger(1957)]{disso} Festinger L (1957). A Theory of
  Cognitive Dissonance.  Stanford University Press, Stanford, CA.


\bibitem{fink_mink}
Fink EL, Kaplowitz SA, Bauer CL (1983). 
Positional discrepancy, psychological discrepancy, and attitude
change: Experimental tests of some mathematical models.
Communication Monographs 50: 413-430.

{
\bibitem{french} French S (1980). Updating of Belief in the Light of
  Someone Else's Opinion.  Journal of the Royal Statistical Society,
  Series A (General): 143 43-48.}


\bibitem{genest_conway}Genest C, McConway KJ (1990). Allocating the
  weights in the linear opinion pool. Journal of Forecasting 9:
  53-73. 

\bibitem[Genest and Zidek(1986)]{genest_zidek} Genest C, Zidek JV
  (1986).  Combining probability distributions: A critique and an
  annotated bibliography. Statistical Science 1: 114--135.

\bibitem[Gentzkow and Shapiro(2006)]{shapiro} Gentzkow M, Shapiro JM
  (2006). Media bias and reputation.  Journal of Political Economy
  114: 280-316.


{
\bibitem{gibbs} Gibbs AL, Su FE (2002). On choosing and bounding
  probability metrics. International statistical review 70:
  419-435.}




\bibitem{hendrick} Hendrick C, Costantini AF (1970). Effects of
  varying trait inconsistency and response requirements on the primacy
  effect in impression formation. Journal of Personality and Social
  Psychology 15: 158-164.

\bibitem[Hogarth and Einhorn(1992)]{Hogarth1992} Hogarth RM, Einhorn
  HJ (1992).  Order effects in belief updating: The belief-adjustment
  model.  Cognitive Psychology, 24: 1-55.

\bibitem{hovland} Hovland CI (editor) (1957). The order of
  presentation in persuasion. Yale University Press, New Haven.

\bibitem{ontology}Huhns MN, Singh MP (1997). Ontologies for agents. IEEE
  Internet Computing. 1: 81-83.




\bibitem{jaynes} Jaynes ET (1968). Prior Probabilities.
  IEEE Trans. Syst. Science \& Cyb.: 4 227-256. 


\bibitem[Jeng(2005)]{jeng} Jeng M (2005).  A selected history of
  expectation bias in physics.  American Journal of Physics, 74:
578-582.


{
\bibitem{sugden}
Jones M, Sugden R (2001). Positive Confirmation Bias In The Acquisition
of Information. Theory and Decision 50: 59-99.}



\bibitem{kaplo}Kaplowitz SA, Fink EL (1997).  Message discrepancy and
  persuasion.  Progress in Communication Sciences XIII. Edited by
  Barnett GA, Foster FJ.  Ablex Publishing Corporation, Greenwhich,
  Connecticut.

{
\bibitem{keller}Keller AM, Winslett M (1985). On the use of extended
  relational model to handle changing incomplete information. IEEE
  Trans. Software Engineering SE-11: 620-633. }



\bibitem{khren} Khrennikov A (2006). Quantum-like brain: Interference
  of minds. BioSystems 84: 225-241.

\bibitem{klayman} Klayman J, Ha YW (1987). Confirmation,
  disconfirmation, and information.  Psychological Review 94: 211-228.

\bibitem[Koehler(1993)]{koehler} Koehler JJ (1993). The influence of
  prior beliefs on scientific judgments of evidence quality.
  Organizational Behavior and Human Decision Processes 56: 28-55.




\bibitem{laroche} Laroche M (1977). A model of attitude change in
  groups following a persuasive communication: An attempt at
  formalizing research findings.  Behavioral Science 22: 246-257.


\bibitem[Lazarsfeld et~al.(1944)Lazarsfeld, Berelson, and
  Gaudet]{lazarsfeld} Lazarsfeld PF, Berelson B, Gaudet H (1944). The
  People's Choice. How the Voter Makes up his Mind in Presidential
  Campaign.  Columbia University Press, New York.


\bibitem{brown} Lindley DV, Tversky A, Brown RV (1979).  On the
  reconciliation of probability assessments.  J. R. Statist. Soc. A
  142: 146-156.


{
\bibitem{lord}
Lord C, Lepper MR, Ross L (1979). Biased Assimilation and Attitude
Polarization: The Effects or Prior Theories on Subsequently Considered
Evidence. Journal of Personality and Social Psychology 37: 2098-2122.}


\comment{
\bibitem[Maher(1993)]{maher}
P. Maher.
\newblock \emph{Betting on Theories}.
\newblock Cambridge University Press, 1993.  }



{
\bibitem{cata} van der Maas HLJ, Kolstein R, van der Pligt J
  (2003). Sudden Transitions in Attitudes. Sociological Methods \&
  Research 32: 125-152.}



\bibitem{miller} Miller N, Campbell DT (1959).  Recency and primacy in
  persuasion as a function of the timing of speeches and measurements.
  The Journal of Abnormal and Social Psychology 59: 1.

\comment{ See also: Neath, I. (1993). Distinctiveness and serial
  position effects in recognition.  Memory and Cognition, 21, 689–698.

Korsnes, M. (1995). Retention intervals and serial list
memory. Perceptual and Motor Skills, 80, 723–731.  }




\bibitem{mulla}
Mullainathan S, Shleifer A (2005). The Market of News.
The American Economic Review 95: 1031-1053.

\bibitem[Nickerson(1998)]{nickerson} Nickerson RS (1998).
  Confirmation bias: a ubiquitous phenomenon in many guises.  Review
  of General Psychology, 2: 175-220.

\bibitem{nyhan} Nyhan B, Reifler J (2010). When corrections fail: The
  persistence of political misperceptions. Political Behavior 32: 303-330.

\bibitem[Oskamp(1965)]{oskamp}
Oskamp (1965).
Overconfidence in case-study judgments.
Journal of Consulting Psychology, 29: 261-265.



\bibitem{peterson}
Peterson CR, DuCharme WM (1967). A primacy effect in
  subjective probability revision. Journal of Experimental
  Psychology 73: 61–65.


\bibitem{pothos} Pothos EM, Busemeyer JR (2009). A quantum probability
  explanation for violations of ``rational'' decision
  theory. Proceedings of the Royal Society B 276: 2171-2178.



\bibitem{rabin}Rabin M, Schrag JL (1999). 
First Impressions Matter: A Model of Confirmatory Bias.
The Quarterly Journal of Economics 114: 37.

\bibitem{boomerang_red} Redlawsk DP, Civettini AJW, Emmerson KM
  (2010).  The Affective Tipping Point: Do Motivated Reasoners Ever
  ``Get It''.  Political Psychology 31: 563-593.

\bibitem{handbook} Social Psychology: Handbook of Basic Principles (2007).
Edited by Kruglanski AW, Higgins EW. The Guilford Press, New York.

\bibitem[Schreider(1970)]{shreider} Schreider YA (1970). On the
  semantic characteristics of information. edited by Saracevic T.
Introduction to Information Science. Bowker, New York: 24-32.

\bibitem[Tversky and Kahneman(1974)]{amos} Tversky A, Kahneman D
  (1974). 
Judgment under uncertainty: Heuristics and biases.
\newblock \emph{Science}, 185\penalty0 (4157):\penalty0 1124-1131.





\bibitem{stewart}
Stewart RH (1965). Effect of continuous responding on
  the order effect in personality impression formation. Journal of
  Personality and Social Psychology 1: 161-165.  

\bibitem{stauffer} Stauffer D (2013). A biased review of
  sociophysics. Journal of Statistical Physics 151: 9-20.



\bibitem{boomerang}Sutherland S (1992). Irrationality: The Enemy
  Within. London, Constable.

{
\bibitem{hungar} Szekely GJ (2005).  Half of a coin: negative
  probabilities. Wilmott Magazine 50: 66-68.}

\bibitem{buse} Trueblood JS, Busemeyer JR (2011). A quantum probability
  account of order effects in inference. Cognitive science 35: 1518-1552.

\bibitem{wason}Wason PC (1960). On the failure to eliminate hypotheses
  in a conceptual task. Quarterly Journal of Experimental Psychology,
  12: 129-140.


\bibitem{krug} Webster DM, Richter L, Kruglanski AW (1996). On leaping
  to conclusions when feeling tired: Mental fatigue effects on
  impressional primacy. Journal of Experimental Social Psychology. 32:
  181-195.

\bibitem[Whittaker(1963)]{Whittaker1963} Whittaker JO (1963).  Opinion
  change as a function of communication-attitude discrepancy.
  Psychological Reports 13: 763-772.


\bibitem{wright} Wright AA, Santiago HC, Sands SF, Kendrick DF, Cook
  RG (1985). Memory processing of serial lists by pigeons, monkeys,
  and people. Science 229: 287-289.

\bibitem{yaniv_advice} Yaniv I (2004). Receiving other people's
  advice: Influence and benefit. Organizational Behavior and Human
  Decision Processes 93: 1–13.

\bibitem{yaniv_trimming} Yaniv I (1997). Weighting and Trimming:
  Heuristics for Aggregating Judgments under
  Uncertainty. Organizational Behavior and Human Decision Processes
  69: 237–249.

\bibitem{yariv}Yariv L (2002). I'll see it when I believe it - A
  Simple Model of Cognitive Consistency. Cowles Foundation Discussion
  Paper \# 1352.






\end{thebibliography}
\end{document}